\documentclass[12pt]{article}

\usepackage{graphicx}
\newcommand{\figref}[1]{Fig.~{\ref{#1}}}
\newcommand{\tabref}[1]{Table~{\ref{#1}}}
\usepackage{amssymb}
\usepackage{amsfonts,amsmath,bm, mathtools}
\usepackage{algorithm}
\usepackage{algorithmic}
\usepackage[numbers,compress]{natbib}
\usepackage{stmaryrd} %jump operator
\usepackage{fancyhdr}

\title{A hybridizable discontinuous Galerkin method for electromagnetics with a view on subsurface applications}

\usepackage{authblk}

\author[1]{Luca Berardocco\footnote{luca.berardocco@tum.de}}
\author[2]{Martin Kronbichler}
\author[3]{Volker Gravemeier\footnote{gravemeier@adco-engineering-gw.com}}
\affil[1]{Continuum Mechanics Group, Technical University of Munich, Boltzmannstr. 15, 85748 Garching, Germany}
\affil[2]{Institute for Computational Mechanics, Technical University of Munich, Boltzmannstr. 15, 85748 Garching, Germany}
\affil[3]{AdCo Engineering$^{GW}$ GmbH, Lichtenbergstr. 8, 85748 Garching, Germany}
\date{}                     %% if you don't need date to appear
\fancypagestyle{titlepagestyle}
{
   \fancyhf{}
   \fancyfoot[C]{\emph{Preprint submitted}}
   
}

\begin{document}

\maketitle
\thispagestyle{titlepagestyle}

\begin{abstract}
Two Hybridizable Discontinuous Galerkin (HDG) schemes for the solution of Maxwell's equations in the time domain are presented. The first method is based on an electromagnetic diffusion equation, while the second is based on Faraday's and Maxwell--Amp\`ere's laws. Both formulations include the diffusive term depending on the conductivity of the medium. The three-dimensional formulation of the electromagnetic diffusion equation in the framework of HDG methods, the introduction of the conduction current term and the choice of the electric field as hybrid variable in a mixed formulation are the key points of the current study. Numerical results are provided for validation purposes and convergence studies of spatial and temporal discretizations are carried out. The test cases include both simulation in dielectric and conductive media.
\end{abstract}

\section{Introduction}
%\begin{itemize}\par
  %\item EM simualtions
  The field of computational electromagnetics has been gaining interest since the first finite difference method for Maxwell's equations was developed by Yee in 1966 \cite{paper.yee}.
  %The method utilizes finite differences to discretise Maxwell's equations in space and therefore implementations of the method are quite efficient.
  With the advent of more powerful computers, it was possible to implement alternative methods such as finite element methods enabling to overcome some of the shortcomings of the original finite difference method, such as problems related to using unstructured meshes and difficulties in achieving higher-order approximations.\par
  The applications of computational electromagnetics are vast and can be found in many engineering fields. Both direct and inverse problems were investigated in the past, e.g., \cite{paper.barucq, paper.haber2004, paper.um10, npc11b, cdl17}, to cite just a few. Examples of fields that benefit from such methods are telecommunications, aeronautics, geophysics, medical imaging and semiconductor industries.\par
  %\item FDTD vs FETD
  Finite elements are generally more computationally expensive than finite differences, but their ability to represent complex geometries and the possibility to locally refine meshes, either via mesh refinement or by increasing the polynomial order of the approximation, makes them very attractive for the present application. The development of such methods greatly benefited from the work of N\'ed\'elec on edge-based elements in \cite{Nedelec1980, Nedelec1986}. Among others, these elements enable a correct representation of the null space of the curl operator; see, e.g.,  \cite{paper.sun95}.\par
  %\item Node Based vs Edge based
  In continuous Galerkin (CG) methods, the continuity between elements is enforced strongly, such that it is therefore not possible to represent discontinuities in the normal component of the electric (or magnetic) field at interfaces between different materials or at domain boundaries. An alternative way to solve this problem, retaining the geometric flexibility and high-order capability, is to use discontinuous Galerkin (DG) methods. In such methods, the continuity between elements is relaxed, such that it is possible to represent discontinuities at element interfaces. However, DG discretizations lead to larger systems of linear equations as compared to CG discretizations.
  %Anyway, thanks to the discontinuous nature of the method, DG implementations can easily be parallelised over many computing units increasing their efficiency in todays computer architectures.
  However, to retain the good properties of DG methods, while addressing their drawbacks, hybridized discontinuous Galerkin (HDG) methods were proposed in \cite{paper.unified_hybridization}. Such methods are obtained from DG methods using flux definitions that reduce the number of globally coupled unknowns and make the solution of the equations faster, as a result. These methods perform similarly to CG methods for matrix-based (see, e.g., \cite{paper.yakovlev2016}) and matrix-free (see, e.g., \cite{paper.kronmatrixfree}) implementations, if appropriate preconditioners are available.
  %A good intoduction to HDG methods can be found in \cite{paper.unified_hybridization}.\par
  %\item Aplication of EM simulations
  %\item Why are we doing this?
  Our research focus on the simulation of electromagnetic diffusion problems in underground strata formations, which might considerably benefit from the advantages of HDG methods. The topic has been investigated using finite differences, e.g., in \cite{paper.wang93, paper.commer2004} and finite elements, e.g., \cite{paper.unsworth93, paper.um10, paper.um12, paper.um14, paper.um15}. While HDG methods were developed for other electromagnetic applications in \cite{npc11b, cdl17}, to the best of our knowledge, they have never been applied to the direct (or inverse) problem of electromagnetic diffusion in geological formations. We focus on an efficient solution of the forward problem, which plays a fundamental role also in inverse problems. To achieve our goal, we first derive and implement a complete HDG formulation for a mixed formulation and validate it by computing test cases which solutions are available in analytical or experimental form. We also derive a formulation based on an electromagnetic diffusion equation that is typically  the equation of choice for geophysical applications.\par
  We would like to point out the differences of our formulation to previous ones proposed in \cite{npc11b, cdl17}. In \cite{npc11b}, a formulation based on the displacement current term was proposed, while in this paper the conduction current is considered. Moreover, while a frequency-domain approach is used in \cite{npc11b}, our approach is formulated in the time domain. A crucial difference between the mixed formulation presented in this paper and the one in \cite{cdl17} is related to the definition of the hybrid variable. In \cite{cdl17}, the hybrid variable is defined as the tangential component of the magnetic field, and an equation is introduced to enforce the electric field to be continuous, while in this paper the opposite choice is made.\par
  %\item Why implicit
  %\item Organization of the paper
  The paper is organised as follow: Section \ref{section.goveq} introduces the constitutive equations in their original form and in the form that will be used as a starting point for our formulation. In Section \ref{section.hdg}, our formulation is derived in a form that is general enough to allow the validation of the method with classical test cases and to obtain the electromagnetic diffusion equation required for subsequent research steps.  Section \ref{section.iti} gives an overview of the chosen implicit time-integration scheme and its implementation. In Section \ref{section.examples}, numerical examples for validating the implementation, including convergence studies, are presented. Finally, Section \ref{section.conclusions} provides conclusions of this work.
%\end{itemize}

\section{Governing equations}
\label{section.goveq}
Classical electrodynamics is mathematically described by electric and magnetic field equations. In combined form, these equations are known as Maxwell's equations. A derivation of the equations as well as a thorough discussion of their mathematical properties can be found, e.g., in \cite{book.jackson99}.

\subsection{Maxwell's equations}
Maxwell's equations are given as follows:
\begin{subequations}\label{eq.mawell}
\begin{align}
\label{faradayslaw}
\nabla\times\mathbf{E} &= -\frac{\partial\mathbf{B}}{\partial t} ,\\[2mm]
\label{maxwellampereslaw}
\nabla\times\mathbf{H} &=\mathbf{J} +\frac{\partial\mathbf{D}}{\partial t},\\[2mm]
\label{elecgausslaw}
\nabla\cdot\mathbf{D} &=\rho ,\\[2mm]
\label{magnegausslaw}
\nabla\cdot\mathbf{B} &= 0.
\end{align}
\end{subequations}
%is known as \textbf{Maxwell's equations}.\\
The variables that appear are: electric field $\mathbf{E}$ ($\mathrm{V/m}$), magnetic field $\mathbf{H}$ ($\mathrm{A/m}$), electric displacement field $\mathbf{D}$ ($\mathrm{C/m^2}$), magnetic induction field $\mathbf{B}$ ($\mathrm{T}$), electric current density $\mathbf{J}$ ($\mathrm{A/m^2}$), and electric charge density $\rho$ ($\mathrm{C/m^3}$). Equations \eqref{faradayslaw}--\eqref{magnegausslaw} are Faraday's law, Maxwell--Amp\`ere's law, (electric) Gauss' law, and (magnetic) Gauss' law, respectively.\par
The equation of continuity, stating conservation of charge, can be used to relate the current $\mathbf{i}$ to the charge density:
\begin{equation}
\label{conteq}
\nabla\cdot\mathbf{i} = -\frac{\partial\rho}{\partial t}
\end{equation}
%only three of five equations \eqref{faradayslaw}-\eqref{conteq} independent: either
%\eqref{faradayslaw}-\eqref{elecgausslaw} or \eqref{faradayslaw}-\eqref{maxwellampereslaw} together with
%\eqref{conteq}
To close the system, it is necessary to add the following constitutive equations, valid for isotropic materials, to equations \eqref{faradayslaw}--\eqref{magnegausslaw}:
\begin{align}
\label{permittivityeq}
\mathbf{D} &= \varepsilon\mathbf{E},\\[2mm]
\label{permeabilityeq}
\mathbf{B} &=\mu\mathbf{H},\\[2mm]
\label{conductivityeq}
\mathbf{J} &=\boldsymbol\sigma\mathbf{E} ,
\end{align}
with the following constitutive parameters: permittivity $\varepsilon$ ($\mathrm{F/m}$), permeability $\mu$
($\mathrm{H/m}$), and conductivity tensor $\boldsymbol\sigma$ ($\mathrm{S/m}$), which is the inverse of the resistivity tensor $\boldsymbol\rho_\mathrm{e}$ ($\Omega\mathrm{m}$).
Equation \eqref{conductivityeq}, which is a generalized form of \textbf{Ohm's law}, may be extended by adding a current-density source $\mathbf{i}_\mathrm{s}$:
\begin{equation}
\label{conductivityeq_ext}
\mathbf{J} =\boldsymbol\sigma\mathbf{E} +\mathbf{i}_\mathrm{s} .
\end{equation}

In this paper, two different boundary conditions are considered. The simplest boundary condition is a Dirichlet-type boundary condition on the tangential component of the electric field, known as \textbf{metallic} or \textbf{perfect electric conductor} (PEC) boundary conditions:
\begin{equation}
\label{bcpce}
\mathbf{n}\times\mathbf{E} =\mathbf{0},
\end{equation}
where $\mathbf{n}$ denotes the outward-pointing unit normal vector to the surface.\par
Another, more complex set of boundary conditions are absorbing boundary conditions (ABC). These boundary conditions are used to truncate physically unbounded domains to finite numerical domains, reducing unwanted reflections. ABC are different in every field of application in which equations enable wave propagation. We derive the ABC for the present problem in the form of the Silver--M\"uller condition according to \cite{paper.barucq} as
\begin{equation}
  \label{eq.silvermueller}
  \sqrt{\varepsilon\mu}\left(\frac{\partial\mathbf{E}}{\partial t}\times\mathbf{n}\right)\times\mathbf{n} - \nabla\times\mathbf{E}\times\mathbf{n} = 0,
\end{equation}
obtaining
\begin{equation}
\label{eq.abc}
\mathbf{H}\times\mathbf{n} +\sqrt{\frac{\varepsilon}{\mu}}\left(\mathbf{E}\times\mathbf{n}\right)\times\mathbf{n}
=\mathbf{H}^\mathrm{inc}\times\mathbf{n}
+\sqrt{\frac{\varepsilon}{\mu}}\left(\mathbf{E}^\mathrm{inc}\times\mathbf{n}\right)\times\mathbf{n}
=\mathbf{g}^\mathrm{inc},
\end{equation}
where $\mathbf{H}$ and $\mathbf{E}$ represent the total fields in the problem domain, while $\mathbf{H}^\mathrm{inc}$ and $\mathbf{E}^\mathrm{inc}$ represent external incoming fields.\par
Furthermore, initial conditions as
\begin{align}
\label{iniconde}
\mathbf{E}_0 &=\mathbf{E}\left(\mathbf{x} , 0\right),\\
\label{inicondh}
\mathbf{H}_0 &=\mathbf{H}\left(\mathbf{x} , 0\right),
\end{align}
are prescribed, such that the initial fields are solutions of the Maxwell system; see, e.g., \cite{paper.rosen80}. To study the response to step-on sources, the initial fields are set to zero everywhere. In geophysical applications, however, where usually the response to step-off sources is of interest, the fields need to be initialized to represent the solution of the static Maxwell equations by solving direct current and magnetometric resistivity problems, as outlined, e.g., in \cite{paper.um10, paper.commer2004, paper.haber2000}. A discussion of such initializations is beyond the scope of this paper, and therefore, in this work, either zero fields or given analytical solutions are used as initial conditions.
%that satisfy the divergence conditions on the fields to avoid erroneous solutions at late times \cite{paper.wang93}.

\subsection{Wave equations for electric and magnetic field}

From equations \eqref{faradayslaw} and \eqref{maxwellampereslaw}, using constitutive equations
\eqref{permittivityeq}, \eqref{permeabilityeq}, and \eqref{conductivityeq_ext}, damped (vector) wave equations for
conductive media can be derived for both the electric field $\mathbf{E}$ and the magnetic flux density
$\mathbf{B}$:
\begin{align}
\label{elewaveq}
\varepsilon\frac{\partial^2\mathbf{E}}{\partial t^2}
+\boldsymbol\sigma\frac{\partial\mathbf{E}}{\partial t}
+\frac{1}{\mu}\nabla\times\nabla\times\mathbf{E} &= -\frac{\partial\mathbf{i}_\mathrm{s}}{\partial t} ,\\
\label{magbwaveq}
\varepsilon\frac{\partial^2\mathbf{B}}{\partial t^2}
+\boldsymbol\sigma\frac{\partial\mathbf{B}}{\partial t}
+\frac{1}{\mu}\nabla\times\nabla\times\mathbf{B} &=\nabla\times\mathbf{i}_\mathrm{s} .
\end{align}
Alternatively, the second wave equation may be formulated in terms of the magnetic field $\mathbf{H}$:
\begin{equation}
\label{maghwaveq}
\varepsilon\frac{\partial^2\mathbf{H}}{\partial t^2}
+\boldsymbol\sigma\frac{\partial\mathbf{H}}{\partial t}
+\frac{1}{\mu}\nabla\times\nabla\times\mathbf{H} =\frac{1}{\mu}\nabla\times\mathbf{i}_\mathrm{s} .
\end{equation}

For subsurface applications, where typically electromagnetic (EM) signals of very low frequency (up to
$f = 10\,\mathrm{Hz}$) are used, the conductivity is usually much larger than the
permittivity (e.g., nine orders of magnitude in rocks); see, e.g., \cite{c10}. As a result, the displacement current becomes negligible with respect to the conduction current. In this case, the damped wave equation reduces to an EM diffusion equation:
\begin{equation}
\label{elediffeq}
\boldsymbol\sigma\frac{\partial\mathbf{E}}{\partial t}
+\frac{1}{\mu}\nabla\times\nabla\times\mathbf{E} = -\frac{\partial\mathbf{i}_\mathrm{s}}{\partial t} .
\end{equation}
The analogous equation for the magnetic field reads
\begin{equation}
\label{magdiffeq}
\boldsymbol\sigma\frac{\partial\mathbf{H}}{\partial t}
+\frac{1}{\mu}\nabla\times\nabla\times\mathbf{H} =\frac{1}{\mu}\nabla\times\mathbf{i}_\mathrm{s} .
\end{equation}
When the frequency goes to zero, the EM diffusion equation \eqref{elediffeq} reduces to the Laplace/Poisson equation
\begin{equation}
\label{lapoieq}
\frac{1}{\mu}\nabla\times\nabla\times\mathbf{E} = -\frac{\partial\mathbf{i}_\mathrm{s}}{\partial t},
\end{equation}
which describes potential-field methods. In this limit, intrinsic resolution becomes almost nonexistent; see \cite{c10}.

In Section \ref{sec.formulation_ele}, the EM diffusion equation \eqref{elediffeq}, along with boundary conditions \eqref{bcpce} and \eqref{eq.abc}, as well as initial condition \eqref{iniconde}, will be considered, and a hybridizable discontinuous Galerkin method will be derived.

\subsection{Mixed equations for electric and magnetic field}

From equations \eqref{faradayslaw} and \eqref{maxwellampereslaw}, using the constitutive equations
\eqref{permittivityeq}, \eqref{permeabilityeq}, and \eqref{conductivityeq_ext}, the following mixed equations
for the electric field $\mathbf{E}$ and the magnetic field $\mathbf{H}$ may also be derived:
\begin{subequations}\label{mixeq}
\begin{align}
\label{mixeq1}
\mu\frac{\partial\mathbf{H}}{\partial t}
+\nabla\times\mathbf{E} &= \mathbf{0} ,\\
\label{mixeq2}
\varepsilon\frac{\partial\mathbf{E}}{\partial t} +\boldsymbol\sigma\mathbf{E}
-\nabla\times\mathbf{H} &= -\mathbf{i}_\mathrm{s} .
\end{align}
\end{subequations}
Assuming the conductivity to be much larger than the permittivity, as above, the equations can be reduced to
\begin{subequations}
\begin{align}
\label{mixeq1fin}
\mu\frac{\partial\mathbf{H}}{\partial t}
+\nabla\times\mathbf{E} &= \mathbf{0} ,\\
\label{mixeq2fin}
\boldsymbol\sigma\mathbf{E}
-\nabla\times\mathbf{H} &= -\mathbf{i}_\mathrm{s} .
\end{align}
\end{subequations}
However, to keep the formulation more general, in Section \ref{sec.formulation_mix}, equations \eqref{mixeq1} and \eqref{mixeq2},
along with boundary conditions \eqref{bcpce} and \eqref{eq.abc} as well as initial
conditions \eqref{iniconde} and \eqref{inicondh}, will be considered, and a hybridizable discontinuous Galerkin
method will be derived.

\section{Hybridizable discontinuous Galerkin method for electromagnetic diffusion equation}
\label{section.hdg}
%Formulation inspired by \cite{npc11a,npc11b,ksmw16,cdl17}.

\subsection{Formulation based on mixed equations}
\label{sec.formulation_mix}
The domain $\Omega$ is discretized with $n_\mathrm{el}$ elements $\Omega_e$ such
that the collection of all elements, $\mathcal{T}_h$, partitions $\Omega$. The boundary of the domain is divided in two disjoint portions, such that $\Gamma_A\cup\Gamma_D=\partial\Omega$. In these two portions, different boundary conditions will be applied, i.e., ABC on $\Gamma_A$ and PEC boundary conditions on $\Gamma_D$. The boundary of the collection, $\partial\mathcal{T}_h$, consists of all element boundaries $\partial\Omega_e$, where
$e = 1, ..., n_\mathrm{el}$. The union of all interior faces $F$, where an interior face denotes the intersection of the
boundaries of two neighboring elements $i$ and $j$ as $F =\partial\Omega_i\cap\partial\Omega_j$, is
denoted by $\mathcal{E}_h^0$. A boundary face is defined as $F =\partial\Omega_i\cap\partial\Omega$, where $i = 1, ..., n_\mathrm{el}$, and the union of such faces is denoted by $\mathcal{E}_h^\partial$. The union of interior and boundary faces is then given by $\mathcal{E}_h=\mathcal{E}_h^\partial\cup\mathcal{E}_h^0$.\par
The space of polynomials of degree of at most $m$ on $D$ is denoted by $\mathcal{P}_m(D)$, where $D$ is an open domain in $\mathbb{R}^d$, and $L^2(D)$ is the space of square integrable-functions on $D$. Furthermore, we set $\mathcal{P}_m^d(D) \equiv [\mathcal{P}_m(D)]^d$. The approximation spaces for solution and weighting functions are then defined as
\begin{align*}
  &\bm{\mathbf{V}}^h = \{\mathbf{v}^{h} \in [L^2(\mathcal{T}_h)]^d : \mathbf{v}^{h}|_{\Omega_e} \in \mathcal{P}_m^d(\Omega_e)\, \forall\Omega_e \in \mathcal{T}_h\},\\
  &\bm{\mathbf{M}}_t^h = \{\boldsymbol{\eta}^{h} \in [L^2(\mathcal{E}_h)]^{d-1} : \boldsymbol{\eta}^{h}|_{F} \in \mathcal{P}_m^{d-1}(\mathcal{E}_h), (\boldsymbol\eta^{h}\cdot\mathbf{n})|_F=0 ~ \forall\Omega_e \in \mathcal{T}_h\},\\
  &\bm{\mathbf{M}}_t^h(\mathbf{f}) = \{\boldsymbol{\eta}^{h} \in \bm{\mathbf{M}}_t^h : \boldsymbol\eta^{h}\times\mathbf{n} = \Pi (\mathbf{f}\times\mathbf{n}) \text{ on } \Gamma_D\}.
\end{align*}
where $\Pi(\mathbf{f}\times\mathbf{n})$ is a projection of $\mathbf{f}\times\mathbf{n}$ onto $\bm{\mathbf{M}}_t^h$. Note that the space $\bm{\mathbf{M}}_t^h$ has a smaller dimension than $\bm{\mathbf{V}}^h$ because the functions in the former are defined on the faces of the elements,  $F$, while the functions in the latter are defined in the elements, $\Omega_e$.\par
Equations \eqref{mixeq1} and \eqref{mixeq2}  are now multiplied by (discrete) weighting functions $(\mathbf{v}^{h}, \mathbf{w}^{h}) \in \bm{\mathbf{V}}^h\times\bm{\mathbf{V}}^h$, respectively, and integrated over one element $\Omega_e$ with boundary $\partial\Omega_e$:
\begin{subequations}
  \begin{align}
    \label{varmixeq1}
    \left(\mathbf{v}^{h} ,\mu\frac{\partial\mathbf{H}^h}{\partial t}\right)_{\Omega_e}
    +\left(\mathbf{v}^{h} ,\nabla\times\mathbf{E}^h\right)_{\Omega_e} &= 0,\\
    \label{varmixeq2}
    \left(\mathbf{w}^{h} ,\varepsilon\frac{\partial\mathbf{E}^h}{\partial t}\right)_{\Omega_e}
    +\left(\mathbf{w}^{h} ,\boldsymbol\sigma\mathbf{E}^h\right)_{\Omega_e}
    -\left(\mathbf{w}^{h} ,\nabla\times\mathbf{H}^h\right)_{\Omega_e}
    &= -\left(\mathbf{w}^{h} ,\mathbf{i}_\mathrm{s}^{h}\right)_{\Omega_e},
  \end{align}
\end{subequations}
where $(\mathbf{E}^h, \mathbf{H}^h)\in \bm{\mathbf{V}}^h\times\bm{\mathbf{V}}^h$ denote approximations of the fields $\mathbf{E} \text{ and } \mathbf{H}$ in the element $\Omega_e$ and $\mathbf{i}_\mathrm{s}^{h}$ is a projection of $\mathbf{i}_\mathrm{s}$ onto $\bm{\mathbf{V}}^h$.\par%The approximation functions $\mathbf{E}^h$ and $\mathbf{H}^h$ belong to the function space $\bm{\mathbf{V}}^h$.
Integration by parts yields
\begin{subequations}
  \begin{align}
    \label{varmixeq1ibp1}
    &\left(\mathbf{v}^{h} ,\mu\frac{\partial\mathbf{H}^h}{\partial t}\right)_{\Omega_e}
    +\left(\nabla\times\mathbf{v}^{h} ,\mathbf{E}^h\right)_{\Omega_e}
    +\left<\mathbf{v}^{h}\times\mathbf{n} ,\hat{\mathbf{E}}^h\right>_{\partial\Omega_e} = 0,\\
    \label{varmixeq2ibp1}
    &\left(\mathbf{w}^{h} ,\varepsilon\frac{\partial\mathbf{E}^h}{\partial t}\right)_{\Omega_e}
    +\left(\mathbf{w}^{h} ,\boldsymbol\sigma\mathbf{E}^h\right)_{\Omega_e}
    -\left(\nabla\times\mathbf{w}^{h} ,\mathbf{H}^h\right)_{\Omega_e}
    -\left<\mathbf{w}^{h}\times\mathbf{n} ,\hat{\mathbf{H}}^h\right>_{\partial\Omega_e}\nonumber\\
    &= -\left(\mathbf{w}^{h} ,\mathbf{i}_\mathrm{s}^{h}\right)_{\Omega_e} ,
  \end{align}
\end{subequations}
where the numerical traces $\hat{\mathbf{E}}^h$ and $\hat{\mathbf{H}}^h$ appear as additional variables.\par%Notice that these additional variables belong to the space $\bm{\mathbf{M}}_t^h$.
As the next step, the hybrid variable $\boldsymbol{\Lambda}^h :=\hat{\mathbf{E}}^h_\mathrm{t}$ is introduced, defined as the tangential component of the numerical trace $\hat{\mathbf{E}}^h$, such that $\boldsymbol{\Lambda}^h \in \bm{\mathbf{M}}_t^h(\mathbf{0})$. Note that the tangential component can be obtained as $\hat{\mathbf{E}}^h_\mathrm{t} =\mathbf{n}\times\left(\hat{\mathbf{E}}^h\times\mathbf{n}\right)$ or as $\hat{\mathbf{E}}^h_\mathrm{t} = \hat{\mathbf{E}}^h -\hat{\mathbf{E}}^h_\mathrm{n}$, where $\hat{\mathbf{E}}^h_\mathrm{n} =\mathbf{n}\left(\hat{\mathbf{E}}^h\cdot\mathbf{n}\right)$. This formualation is similar to the one in \cite{cdl17}, with the difference that here the hybrid variable is defined as the tangential component of the electric instead of the magnetic field. The main motivation for this choice lies in the fact that the tangential component of the electric field is always continuous. While charge density, surface currents and material heterogeneity can induce discontinuities in any component of the magnetic field or in the normal component of the electric field, $\mathbf{E}_t$ is always continuous, as emphasized in \cite{book.jackson99} (Chapter 4--5). There is not any notable difference in dielectric media, where surface currents are not allowed, but there might be one in conductive media, such as the ones encountered in geophysical applications.\par
%\textcolor{red}{The motivation to use the electric, instead of magnetic, field lies on the field of application of the method. For electromagnetic diffusion in conductive media it is possible to have surface currents in the domain that would create discontinuities in the magnetic, but not in the electric, field.}\par
Summing all contributions by individual elements, the following equation is obtained:
\begin{subequations}\label{eq.varmix}
  \begin{align}
    \label{varmixeq1ibp}
    &\left(\mathbf{v}^{h} ,\mu\frac{\partial\mathbf{H}^{h}}{\partial t}\right)_{\mathcal{T}^h}
    +\left(\nabla\times\mathbf{v}^{h} ,\mathbf{E}^h\right)_{\mathcal{T}^h}
    +\left<\mathbf{v}^{h}\times\mathbf{n} ,\boldsymbol{\Lambda}^h\right>_{\partial\mathcal{T}^h} = 0,\\
    \label{varmixeq2ibp}
    &\left(\mathbf{w}^{h} ,\varepsilon\frac{\partial\mathbf{E}^h}{\partial t}\right)_{\mathcal{T}^h}
    +\left(\mathbf{w}^{h} ,\boldsymbol\sigma\mathbf{E}^h\right)_{\mathcal{T}^h}
    -\left(\nabla\times\mathbf{w}^{h} ,\mathbf{H}^h\right)_{\mathcal{T}^h}
    -\left<\mathbf{w}^{h}\times\mathbf{n} ,\hat{\mathbf{H}}^h\right>_{\partial\mathcal{T}^h}\nonumber\\
    &= -\left(\mathbf{w}^{h} ,\mathbf{i}_\mathrm{s}^{h}\right)_{\mathcal{T}^h} .
  \end{align}
\end{subequations}
The numerical trace of the magnetic field $\hat{\mathbf{H}}^h$ is defined
here as a sum of $\mathbf{H}^h$ and a stabilization term, weighting the difference between the
tangential component of the discrete electrical field $\mathbf{E}^h$ and the hybrid variable
$\boldsymbol{\Lambda}^h$:
\begin{equation}\label{eq.htracedef}
\hat{\mathbf{H}}^h =\mathbf{H}^h
+\tau\left(\mathbf{E}^h_\mathrm{t} -\boldsymbol{\Lambda}^h\right)\times\mathbf{n} ,
\end{equation}
where $\tau$ is a stabilization parameter.\par
%where the stabilization parameter is defined as
%\begin{equation}
%\tau =\sqrt{\frac{\left|\boldsymbol\sigma\right|}{\mu\, t_\mathrm{c}}} ,
%\end{equation}
%where $t_\mathrm{c}$ is a characteristic time scale.
Via an additional equation, the continuity of the tangential component of $\hat{\mathbf{H}}^h$
across inter-element boundaries is enforced:
\begin{equation}\label{eq.jumph}
\llbracket\mathbf{n}\times\hat{\mathbf{H}}^h\rrbracket = 0\quad\mathrm{on}\;\mathcal{E}_h^0 ,
\end{equation}
with the definition
\begin{equation*}
  \llbracket\mathbf{n}\times\hat{\mathbf{H}}^h\rrbracket := \mathbf{n}^+\times\hat{\mathbf{H}}^{h+} + \mathbf{n}^-\times\hat{\mathbf{H}}^{h-}
\end{equation*}
for interior faces, or simply
\begin{equation*}
  \llbracket\mathbf{n}\times\hat{\mathbf{H}}^h\rrbracket := \mathbf{n}\times\hat{\mathbf{H}}^{h}
\end{equation*}
for boundary faces.\par
Definition \eqref{eq.htracedef} and condition \eqref{eq.jumph}, are required to close the system of equations and ensuring the numerical trace $\hat{\mathbf{H}}^h$ to be single-valued. As a result, the method is locally conservative. To prove that \eqref{eq.jumph} ensures a single-valued numerical trace $\hat{\mathbf{H}}^h$, equation \eqref{eq.htracedef} is inserted in \eqref{eq.jumph}, yielding
\begin{equation}
  \llbracket\mathbf{n}\times\mathbf{H}^h\rrbracket + \tau^+\mathbf{E}^{h+}_\mathrm{t} + \tau^-\mathbf{E}^{h-}_\mathrm{t} - (\tau^+ + \tau^-)\boldsymbol{\Lambda}^h = 0,
\end{equation}
from which it is possible to obtain an equation for the hybrid variable:
\begin{equation}\label{eq.hybridvalue}
  \boldsymbol{\Lambda}^h = \frac{\tau^+\mathbf{E}^{h+}_\mathrm{t} + \tau^-\mathbf{E}^{h-}_\mathrm{t}}{\tau^+ + \tau^-} + \frac{1}{\tau^+ + \tau^-}\llbracket\mathbf{n}\times\mathbf{H}^h\rrbracket.
\end{equation}
Substituting \eqref{eq.hybridvalue} in \eqref{eq.htracedef}, the following equation is obtained
\begin{equation}
  \hat{\mathbf{H}}^h =\mathbf{H}^h +\tau\left(\mathbf{E}^h_\mathrm{t} - \frac{\tau^+\mathbf{E}^{h+}_\mathrm{t} + \tau^-\mathbf{E}^{h-}_\mathrm{t}}{\tau^+ + \tau^-} - \frac{1}{\tau^+ + \tau^-}\llbracket\mathbf{n}\times\mathbf{H}^h\rrbracket\right)\times\mathbf{n},
\end{equation}
eventually yielding
\begin{equation}\label{eq.uniquenessproof}
  \hat{\mathbf{H}}^{h-} = \hat{\mathbf{H}}^{h+} = \hat{\mathbf{H}}^h = \frac{\tau^-\mathbf{H}^{h+}_\mathrm{t} + \tau^+\mathbf{H}^{h-}_\mathrm{t}}{\tau^+ + \tau^-} + \frac{\tau^+\tau^-}{\tau^+ + \tau^-}\llbracket\mathbf{E}^h\times\mathbf{n}\rrbracket.
\end{equation}
Note that to obtain \eqref{eq.uniquenessproof}, the relation $\mathbf{n}^+ = -\mathbf{n}^-$ is used.\par
For an individual element boundary $\partial\Omega_e$, the variational form of \eqref{eq.jumph}, with weighting function $\boldsymbol{\eta}^{h}$, is given as
\begin{equation}\label{eq.elementcontinuitymix}
\left<\boldsymbol{\eta}^{h} ,\mathbf{n}\times\hat{\mathbf{H}}^h\right>_{\partial\Omega_e} = 0 .
\end{equation}
To include the ABCs defined by \eqref{eq.abc}, equation \eqref{eq.elementcontinuitymix} is integrated over the whole domain and the boundary condition terms are added as
\begin{equation}\label{eq.continuity}
  -\langle\boldsymbol{\eta}^{h}, \hat{\mathbf{H}}^h\times\mathbf{n}\rangle_{\partial\tau_h} + \left\langle\boldsymbol{\eta}^{h}, \sqrt{\frac{\varepsilon}{\mu}}\boldsymbol{\Lambda}^{h}\right\rangle_{\Gamma_A} = -\left\langle\boldsymbol{\eta}^{h}, \mathbf{g}^h_{\mathrm{inc}}\right\rangle_{\Gamma_A},
\end{equation}
where $\mathbf{g}^h_{\mathrm{inc}}$ is a projection of $\mathbf{g}_{\mathrm{inc}}$ onto $\mathbf{M}_t^h$.\par
Adding \eqref{eq.continuity} to equations \eqref{eq.varmix} yields%the following system of equations needs to be solved: find $\left(\mathbf{H}^h ,\mathbf{E}^h ,\boldsymbol{\Lambda}^h\right)$ such that
\begin{subequations}
  \begin{align}
    \label{hdgmixeq1}
    &\left(\mathbf{v}^{h} ,\mu\frac{\partial\mathbf{H}^{h}}{\partial t}\right)_{\mathcal{T}^h}
    +\left(\nabla\times\mathbf{v}^{h} ,\mathbf{E}^h\right)_{\mathcal{T}^h}
    +\left<\mathbf{v}^{h}\times\mathbf{n} ,\boldsymbol{\Lambda}^h\right>_{\partial\mathcal{T}^h} = 0,\\[2mm]
    \label{hdgmixeq2}
    &\left(\mathbf{w}^{h} ,\varepsilon\frac{\partial\mathbf{E}^h}{\partial t}\right)_{\mathcal{T}^h}
    +\left(\mathbf{w}^{h} ,\boldsymbol\sigma\mathbf{E}^h\right)_{\mathcal{T}^h}
    -\left(\nabla\times\mathbf{w}^{h} ,\mathbf{H}^h\right)_{\mathcal{T}^h}\nonumber\\[2mm]
    &-\left<\mathbf{w}^{h}\times\mathbf{n} ,\mathbf{H}^h\right>_{\partial\mathcal{T}^h}
    -\left<\mathbf{w}^{h}\times\mathbf{n} ,\tau\left(\mathbf{E}^h_\mathrm{t}
    -\boldsymbol{\Lambda}^h\right)\times\mathbf{n}\right>_{\partial\mathcal{T}^h}
    = -\left(\mathbf{w}^{h} ,\mathbf{i}_\mathrm{s}^{h}\right)_{\mathcal{T}^h} ,\\[2mm]
    \label{hdgmixeq3}
    &-\left<\boldsymbol{\eta}^{h} ,\mathbf{H}^h\times\mathbf{n}\right>_{\partial\mathcal{T}^h}
    +\left<\boldsymbol{\eta}^{h} ,\mathbf{n}\times\left(\tau\left(\mathbf{E}^h_\mathrm{t}
    -\boldsymbol{\Lambda}^h\right)\times\mathbf{n}\right)\right>_{\partial\mathcal{T}^h}
    +\left\langle\boldsymbol{\eta}^{h}, \sqrt{\frac{\varepsilon}{\mu}}\boldsymbol{\Lambda}^{h}\right\rangle_{\Gamma_A}\nonumber\\[2mm]
    &= -\left\langle\boldsymbol{\eta}^{h},   \mathbf{g}_{\mathrm{inc}}^{h}\right\rangle_{\Gamma_A}.
  \end{align}
\end{subequations}
%for all $\left(\mathbf{v}^{h} ,\mathbf{w}^{h} ,\boldsymbol{\eta}^{h}^h\right)$, which might be rewritten as
Applying integration by parts to \eqref{hdgmixeq2} and rearranging the equations leads to the final problem: find $\left(\mathbf{H}^h ,\mathbf{E}^h ,\boldsymbol{\Lambda}^h\right)\in\bm{\mathbf{V}}^h\times\bm{\mathbf{V}}^h\times\bm{\mathbf{M}}^h(0)$ such that
\begin{subequations}\label{eq.hdgmixfinal}
  \begin{align}
    \label{hdgmixeq1final}
    &\left(\mathbf{v}^{h} ,\mu\frac{\partial\mathbf{H}^{h}}{\partial t}\right)_{\mathcal{T}^h}
    +\left(\nabla\times\mathbf{v}^{h} ,\mathbf{E}^h\right)_{\mathcal{T}^h}
    +\left<\mathbf{v}^{h}\times\mathbf{n} ,\boldsymbol{\Lambda}^h\right>_{\partial\mathcal{T}^h} = 0,\\[2mm]
    \label{hdgmixeq2final}
    &\left(\mathbf{w}^{h} ,\varepsilon\frac{\partial\mathbf{E}^h}{\partial t}\right)_{\mathcal{T}^h}
    +\left(\mathbf{w}^{h} ,\boldsymbol\sigma\mathbf{E}^h\right)_{\mathcal{T}^h}
    -\left(\mathbf{w}^{h} ,\nabla\times\mathbf{H}^h\right)_{\mathcal{T}^h}
    -\left<\mathbf{w}^{h}\times\mathbf{n} ,\tau\mathbf{E}^h_\mathrm{t}\times\mathbf{n}\right>_{\partial\mathcal{T}^h}\nonumber\\[2mm]
    &+\left<\mathbf{w}^{h}\times\mathbf{n} ,\tau\boldsymbol{\Lambda}^h\times\mathbf{n}\right>_{\partial\mathcal{T}^h}
    = -\left(\mathbf{w}^{h} ,\mathbf{i}_\mathrm{s}^{h}\right)_{\mathcal{T}^h} ,\\[2mm]
    \label{hdgmixeq3final}
    &-\left<\boldsymbol{\eta}^{h} ,\mathbf{H}^h\times\mathbf{n}\right>_{\partial\mathcal{T}^h}
    +\left<\boldsymbol{\eta}^{h} ,\tau\mathbf{E}^h_\mathrm{t}\right>_{\partial\mathcal{T}^h}
    -\left<\boldsymbol{\eta}^{h} ,\tau\boldsymbol{\Lambda}^h\right>_{\partial\mathcal{T}^h}
    +\left\langle\boldsymbol{\eta}^{h}, \sqrt{\frac{\varepsilon}{\mu}}\boldsymbol{\Lambda}^{h}\right\rangle_{\Gamma_A}\nonumber\\[2mm]
    &= -\left\langle\boldsymbol{\eta}^{h}, \mathbf{g}_{\mathrm{inc}}^{h}\right\rangle_{\Gamma_A}.
  \end{align}
\end{subequations}
for all $\left(\mathbf{v}^{h} ,\mathbf{w}^{h} ,\boldsymbol{\eta}^{h}\right)\in\bm{\mathbf{V}}^h\times\bm{\mathbf{V}}^h\times\bm{\mathbf{M}}^h(0)$.\\
From these equations, the following matrix-vector formulation is obtained:
\begin{subequations}
\label{semimatrixmix}
  \begin{equation}
   \begin{bmatrix}
   \mathbb{A}^{\mathrm{m}} & \mathbf{0} \\
   \mathbf{0} & \mathbb{E}^{\mathrm{m}}
   \end{bmatrix}
   \begin{bmatrix}
   \dot{\mathbf{H}} \\
   \dot{\mathbf{E}}
   \end{bmatrix}
   +
   \begin{bmatrix}
   \mathbf{0} & \mathbb{C}^{\mathrm{m}} \\
   \mathbb{F}^{\mathrm{m}} & \mathbb{G}^{\mathrm{m}}
   \end{bmatrix}
   \begin{bmatrix}
   \mathbf{H}  \\
   \mathbf{E}
   \end{bmatrix}
   +
   \begin{bmatrix}
   \mathbb{D}^{\mathrm{m}} \\
   \mathbb{H}^{\mathrm{m}}
   \end{bmatrix}
   \boldsymbol\Lambda
   =
   \begin{bmatrix}
   \mathbf{0}   \\
   -\mathbf{I}_\mathrm{s}
   \end{bmatrix},
  \end{equation}
    \begin{equation}
      \mathbb{I}^{\mathrm{m}}\mathbf{H} +\mathbb{J}^{\mathrm{m}}\mathbf{E} + \mathbb{L}^{\mathrm{m}}\boldsymbol\Lambda = 0.
    \end{equation}
\end{subequations}

\subsection{Formulation based on equations for electric field}
\label{sec.formulation_ele}
The derivation of the second formulation is sketched here, and the final form is given. The complete derivation closely follows the steps outlined above for the mixed equations.\par
A new variable $\mathbf{u} =\mu^{-1}\nabla\times\mathbf{E}$ is introduced into the EM diffusion equation \eqref{elediffeq}, such that it can be split up into two equations as follows:
of equations is obtained:
\begin{subequations}\label{elewave}
  \begin{align}
    \label{elewaveq1}
    \mu\mathbf{u} -\nabla\times\mathbf{E} &= 0,\\
    \label{elewaveq2}
    \boldsymbol\sigma\frac{\partial\mathbf{E}}{\partial t}
    +\nabla\times\mathbf{u} &= -\frac{\partial\mathbf{i}_\mathrm{s}}{\partial t} ,
  \end{align}
\end{subequations}
with PEC boundary conditions \eqref{bcpce}.
The new variable $\mathbf{u}$ is related to the magnetic field $\mathbf{H}$ via equations \eqref{faradayslaw} and \eqref{permeabilityeq} as follows:
\begin{equation}
\label{htou}
\mathbf{u} = -\frac{\partial\mathbf{H}}{\partial t} ,
\end{equation}
and vice versa,
\begin{equation}
\mathbf{H} = -\int_t\mathbf{u}\,\mathrm{d} t .
\end{equation}
Note the differences of the present system of equations \eqref{elewave} to \eqref{mixeq}, wich was the starting point for the derivation in the preceding section.\par
Based on the discretization already introduced above, the equations are multiplied by (discrete) weighting functions $(\mathbf{v}^{h}, \mathbf{w}^{h}) \in \bm{\mathbf{V}}^h\times\bm{\mathbf{V}}^h$, respectively, and integrated over one element $\Omega_e$ with boundary $\partial\Omega_e$:
\begin{subequations}
  \begin{align}
    \label{varelewaveq1}
    \left(\mathbf{v}^{h} ,\mu\mathbf{u}^h\right)_{\Omega_e}
    -\left(\mathbf{v}^{h} ,\nabla\times\mathbf{E}^h\right)_{\Omega_e} &= 0,\\
    \label{varelewaveq2}
    \left(\mathbf{w}^{h} ,\boldsymbol\sigma\frac{\partial\mathbf{E}^h}{\partial t}\right)_{\Omega_e}
    +\left(\mathbf{w}^{h} ,\nabla\times\mathbf{u}^h\right)_{\Omega_e}
    &= -\left(\mathbf{w}^{h} ,\frac{\partial\mathbf{i}_\mathrm{s}^{h}}{\partial t}\right)_{\Omega_e}.
  \end{align}
\end{subequations}
where $(\mathbf{u}^h, \mathbf{E}^h)\in \bm{\mathbf{V}}^h\times\bm{\mathbf{V}}^h$ are the approximations of the fields $\mathbf{u}$ and $\mathbf{E}$, respectively, and $\mathbf{i}_\mathrm{s}^{h}$ is a projection of $\mathbf{i}_\mathrm{s}$ onto $\bm{\mathbf{V}}^h$.\par
Summing all contributions by individual elements, integrating by parts and substituting the definitions
\begin{align*}
  &\boldsymbol\Lambda^h:=\hat{\mathbf{E}}^h_t,\\
  &\hat{\mathbf{u}}^h =\mathbf{u}^h +\tau\left(\mathbf{E}^h_\mathrm{t} -\boldsymbol{\Lambda}^h\right)\times\mathbf{n},
\end{align*}
yields
\begin{subequations}
  \begin{align}
    \label{varelewaveq1final}
    \left(\mathbf{v}^{h} ,\mu\mathbf{u}^h\right)_{\mathcal{T}^h}
    -&\left(\nabla\times\mathbf{v}^{h} ,\mathbf{E}^h\right)_{\mathcal{T}^h}
    -\left<\mathbf{v}^{h}\times\mathbf{n} ,\boldsymbol{\Lambda}^h\right>_{\partial\mathcal{T}^h} = 0,\\[2mm]
    \label{varelewaveq2final}
    \left(\mathbf{w}^{h} ,\boldsymbol\sigma\frac{\partial\mathbf{E}^h}{\partial t}\right)_{\mathcal{T}^h}
    &+\left(\nabla\times\mathbf{w}^{h} ,\mathbf{u}^h\right)_{\mathcal{T}^h}
    +\left<\mathbf{w}^{h}, \mathbf{n}\times\mathbf{u}^h\right>_{\partial\mathcal{T}^h}\nonumber\\%[2mm]
    &+\left<\mathbf{w}^{h},\tau\left(\mathbf{E}^h_\mathrm{t}
    -\boldsymbol{\Lambda}^h\right)\right>_{\partial\mathcal{T}^h}
    = -\left(\mathbf{w}^{h} ,\frac{\partial\mathbf{i}_\mathrm{s}^{h}}{\partial t}\right)_{\mathcal{T}^h}.
  \end{align}
\end{subequations}

To close the system of equations the continuity equation
\begin{equation}
  \left<\boldsymbol{\eta}^{h} ,\mathbf{n}\times\mathbf{u}^h\right>_{\partial\mathcal{T}^h}
  +\left<\boldsymbol{\eta}^{h} ,\mathbf{n}\times\left(\tau\left(\mathbf{E}^h_\mathrm{t}
  -\boldsymbol{\Lambda}^h\right)\times\mathbf{n}\right)\right>_{\partial\mathcal{T}^h} = 0
\end{equation}
is added.\par
The final problem reads: find $\left(\mathbf{u}^h ,\mathbf{E}^h ,\boldsymbol{\Lambda}^h\right)\in\mathbf{V}^h\times\mathbf{V}^h\times\mathbf{M}^h_t(0)$ such that
\begin{subequations}
  \begin{align}
    \label{hdgelewaveq1final}
    \left(\mathbf{v}^{h} ,\mu\mathbf{u}^h\right)_{\mathcal{T}^h}
    -&\left(\nabla\times\mathbf{v}^{h} ,\mathbf{E}^h\right)_{\mathcal{T}^h}
    -\left<\mathbf{v}^{h}\times\mathbf{n} ,\boldsymbol{\Lambda}^h\right>_{\partial\mathcal{T}^h} = 0,\\[2mm]
    \label{hdgelewaveq2final}
    \left(\mathbf{w}^{h} ,\boldsymbol\sigma\frac{\partial\mathbf{E}^h}{\partial t}\right)_{\mathcal{T}^h}
    &+\left(\mathbf{w}^{h} ,\nabla\times\mathbf{u}^h\right)_{\mathcal{T}^h}
    +\left<\mathbf{w}^{h},\tau\mathbf{E}^h_\mathrm{t}\right>_{\partial\mathcal{T}^h}\nonumber\\[2mm]
    &-\left<\mathbf{w}^{h},\tau\boldsymbol{\Lambda}^h\right>_{\partial\mathcal{T}^h}
    =-\left(\mathbf{w}^{h} ,\frac{\partial\mathbf{i}_\mathrm{s}^{h}}{\partial t}\right)_{\mathcal{T}^h} ,\\[2mm]
    \label{hdgelewaveq3final}
    -\left<\boldsymbol{\eta}^{h} ,\mathbf{u}^h\times\mathbf{n}\right>_{\partial\mathcal{T}^h}
    &+\left<\boldsymbol{\eta}^{h} ,\tau\mathbf{E}^h_\mathrm{t}\right>_{\partial\mathcal{T}^h}
    -\left<\boldsymbol{\eta}^{h} ,\tau\boldsymbol{\Lambda}^h\right>_{\partial\mathcal{T}^h}
    = 0,
  \end{align}
\end{subequations}
for all $\left(\mathbf{v}^{h} ,\mathbf{w}^{h} ,\boldsymbol{\eta}^{h}\right)\in\mathbf{V}^h\times\mathbf{V}^h\times\mathbf{M}^h_t(0)$.\\
From this system of equations, the following matrix-vector formulation is obtained:
\begin{subequations}
\label{semimatrixwave}
  \begin{equation}
   \begin{bmatrix}
   \mathbf{0} & \mathbf{0}\\
   \mathbf{0} & \mathbb{E}^{\mathrm{e}}
   \end{bmatrix}
   \begin{bmatrix}
   \dot{\mathbf{U}}\\
   \dot{\mathbf{E}}
   \end{bmatrix}
   +
   \begin{bmatrix}
   \mathbb{A}^{\mathrm{e}} & \mathbb{B}^{\mathrm{e}} \\
   \mathbb{F}^{\mathrm{e}} & \mathbb{G}^{\mathrm{e}}
   \end{bmatrix}
   \begin{bmatrix}
   \mathbf{U}  \\
   \mathbf{E}
   \end{bmatrix}
   +
   \begin{bmatrix}
   \mathbb{C}^{\mathrm{e}} \\
   \mathbb{H}^{\mathrm{e}}
   \end{bmatrix}
   \boldsymbol\Lambda
   =
   \begin{bmatrix}
   \mathbf{0}   \\
   -\dot{\mathbf{I}}_\mathrm{s}
   \end{bmatrix}   ,
\end{equation}
\begin{equation}
\mathbb{I}^{\mathrm{e}}\mathbf{U} +\mathbb{J}^{\mathrm{e}}\mathbf{E} + \mathbb{L}^{\mathrm{e}}\boldsymbol\Lambda = 0.
\end{equation}
\end{subequations}
Note that, even if the same notation is used, apart from $\mathbb{I} \text { and } \mathbb{J}$, the matrices in \eqref{semimatrixwave} are different from the ones in \eqref{semimatrixmix}. Note further the different structure of the matrices in that a zero matrix apperas in the upper left block for the non-time-dependent $(\mathbf{H}, \mathbf{E})$ $2\times2$ system in \eqref{semimatrixmix}, while it is observed for the time-dependent $(\mathbf{\dot{U}}, \mathbf{\dot{E}})$ $2\times2$ system in \eqref{semimatrixwave}.

\section{Implicit time-integration}
\label{section.iti}
Due to the frequency content and high variability of the constitutive parameters of the target application, the equation is very stiff in  time \cite{paper.haber2004}. Therefore, in our implementation, an implicit time-integration is chosen, to avoid time-step restrictions of explicit time-integration schemes. In this section, the implicit time-integration for the formulation in Section \ref{sec.formulation_mix} is shown, by using a backward-Euler scheme for demonstration purposes. Previous matrix and vector indices are neglected to simplify the notation. The implementation of a higher-order time-integration scheme would follow a similar rationale.\par
Applying the backward-Euler scheme to \eqref{semimatrixmix}, the following matrix-vector system is obtained
\begin{gather}
   \label{matrixmix}
   \begin{bmatrix}
   \frac{1}{\Delta t}\mathbb{A} & \mathbb{C} & \mathbb{D} \\
   \mathbb{F} & \frac{1}{\Delta t}\mathbb{E} +\mathbb{G} & \mathbb{H} \\
   \mathbb{I} & \mathbb{J} & \mathbb{L}
   \end{bmatrix}
   \begin{bmatrix}
   \mathbf{H}^{n + 1} \\
   \mathbf{E}^{n + 1} \\
   \boldsymbol\Lambda^{n + 1}
   \end{bmatrix}
   =
   \begin{bmatrix}
   \frac{1}{\Delta t}\mathbb{A}\mathbf{H}^n   \\
   \frac{1}{\Delta t}\mathbb{E}\mathbf{E}^n -\mathbf{I}_\mathrm{s}^{n+1} \\
   \mathbf{0}
   \end{bmatrix},
\end{gather}
where $\Delta t$ is the chosen time-step and the superscript $n$ indicates that we are referring to the solution at the time $t_0 + n\Delta t$.\par
Condensation of the upper left $2\times2$ block yields a system of equations for the degrees of freedom of the hybrid variable exclusively:
\begin{equation}
\label{eq.condensedmix}
\mathbb{K}\boldsymbol\Lambda^{n+1} =\mathbf{F} ,
\end{equation}
with
\begin{gather}
   \mathbb{K} =
   \mathbb{L} -
   \begin{bmatrix}
   \mathbb{I} & \mathbb{J}
   \end{bmatrix}
   \begin{bmatrix}
   \frac{1}{\Delta t}\mathbb{A} & \mathbb{C} \\
   \mathbb{F} & \frac{1}{\Delta t}\mathbb{E} +\mathbb{G}
   \end{bmatrix}^{- 1}
   \begin{bmatrix}
   \mathbb{D} \\
   \mathbb{H}
   \end{bmatrix}
\end{gather}
and
\begin{gather}
   \mathbf{F} = -
   \begin{bmatrix}
   \mathbb{I} & \mathbb{J}
   \end{bmatrix}
   \begin{bmatrix}
   \frac{1}{\Delta t}\mathbb{A} & \mathbb{C} \\
   \mathbb{F} & \frac{1}{\Delta t}\mathbb{E} +\mathbb{G}
   \end{bmatrix}^{- 1}
   \begin{bmatrix}
   \frac{1}{\Delta t}\mathbb{A}\mathbf{H}^n   \\
   \frac{1}{\Delta t}\mathbb{E}\mathbf{E}^n -\mathbf{I}_\mathrm{s}^{n+1}
   \end{bmatrix}.
\end{gather}
After having solved \eqref{eq.condensedmix}, the interior degrees of freedom of the magnetic and electric fields are computed elementwise as
\begin{gather}
   \begin{bmatrix}
   \mathbf{H}^{n + 1} \\
   \mathbf{E}^{n + 1}
   \end{bmatrix}
    =
   \begin{bmatrix}
   \frac{1}{\Delta t}\mathbb{A} & \mathbb{C} \\
   \mathbb{F} & \frac{1}{\Delta t}\mathbb{E} +\mathbb{G}
   \end{bmatrix}^{- 1}
   \left(
   \begin{bmatrix}
   \frac{1}{\Delta t}\mathbb{A}\mathbf{H}^n   \\
   \frac{1}{\Delta t}\mathbb{E}\mathbf{E}^n -\mathbf{I}_\mathrm{s}^{n+1}
   \end{bmatrix} -
   \begin{bmatrix}
   \mathbb{D} \\
   \mathbb{H}
   \end{bmatrix}
   \boldsymbol\Lambda^{n + 1}\right) .
\end{gather}
The time-discretized formulation according to Section \ref{sec.formulation_ele} can be obtained in a similar way, but the derivation is here omitted for brevity.

\section{Numerical examples}
\label{section.examples}
In this section, four numerical examples demonstrate that our method is able to reproduce physical phenomena such as wave propagation and scattering correctly as well as the possibility to utilize unstructured meshes and absorbing boundary conditions to bound numerical domains. The chosen implicit time-integration scheme makes the solution of a system of linear equations necessary for every time step of the simulation. There are several approaches that can be used to solve a system of linear equations, which efficiency depends upon the properties of the system itself. In this study, we use SuperLU\_DIST \cite{report.superlu99, paper.li03} within Amesos interface \cite{paper.heroux05}, which is sufficient for the size of the systems of linear equations considered in the following four examples. For larger problem sizes, iterative solvers will become necessary. However, the development of specific solvers and preconditioners, respectively, for the problems such as the present one will be the subject of future work.

\subsection{Cubic cavity with perfect electric conductor boundary conditions}
\label{subsection.cubicpec}
In this subsection, the solution of Maxwell's equations in a cubic cavity with perfectly conductive boundary conditions is presented. The domain is a unit cube, discretized with hexahedral elements. The permittivity and permeability are set to $\varepsilon = \sqrt{3}\,\text{F/m}$ and $\mu = \sqrt{3}\,\text{H/m}$, respectively.\par
The frequency of the resonant wave is given by% \cite{book.pozar}
\begin{equation}
  f_{mnp} = \frac{1}{2\sqrt{\mu\varepsilon}}\left[\left(\frac{m}{a}\right)^2+\left(\frac{n}{b}\right)^2+\left(\frac{p}{d}\right)^2\right]^{\frac{1}{2}},
\end{equation}
where $m, n$ and $p$ are integers that define the wave pattern in the domain, and $a, b$ and $d$ are the dimensions of the cavity. The period of the wave can then be obtained using the relation $T_{mnp} = 1/f_{mnp}$, which is $T_{111} = 2\,$s in this test case.\par
The analytical solution for this mode reads
\begin{equation}
  \label{eq.initial}
  \begin{split}
  \mathbf{E}^{\mathrm{a}} =& \left\{
  \begin{matrix*}[l]
    E^{\mathrm{a}}_x = -\sin(\pi x) \cos(\pi y) \cos(\pi z) \cos(\pi t),\\[2mm]
    E^{\mathrm{a}}_y = 2 \cos(\pi x) \sin(\pi y) \cos(\pi z) \cos(\pi t),\\[2mm]
    E^{\mathrm{a}}_z = -\cos(\pi x) \cos(\pi y) \sin(\pi z) \cos(\pi t),
  \end{matrix*}\right.\\[2mm]
  \mathbf{H}^{\mathrm{a}} =& \left\{
  \begin{matrix*}[l]
    H^{\mathrm{a}}_x = -\sqrt{3} \cos(\pi x) \sin(\pi y) \sin(\pi z) \sin(\pi t),\\[2mm]
    H^{\mathrm{a}}_y = 0,\\[2mm]
    H^{\mathrm{a}}_z = \sqrt{3} \sin(\pi x) \sin(\pi y) \cos(\pi z) \sin(\pi t).
  \end{matrix*}\right.
\end{split}
\end{equation}
The intial conditions are obtained by \eqref{eq.initial}, when setting $t = 0$.\par
The error considered here is a relative error computed on the electric field in the $L_2$-norm, which is defined as
\begin{equation}
  \|\mathbf{E}^{\mathrm{a}} - \mathbf{E}_h\|_{\tau_h} =  \left(\sum_{\tau_h}\frac{\int_{\Omega_e}\|\mathbf{E}^{\mathrm{a}} - \mathbf{E}_h\|^2}{\int_{\Omega_e}\|\mathbf{E}^{\mathrm{a}}\|^2}\right)^{1/2}.
\end{equation}
\subsubsection{Spatial convergence}\label{subsubsection.spatial_convergence}
To make sure that the two sources of errors resulting from spatial and temporal discretization do not interfere with each other, we study the spatial convergence rates for a particularly very small time step. The chosen time-integration scheme is only of first-order accuracy and therefore small time steps are anyway recomended. The time-step length chosen here is $\Delta t = 10^{-4}\,$s.
\begin{figure}[ht]
\centering
\includegraphics[width=.95\textwidth]{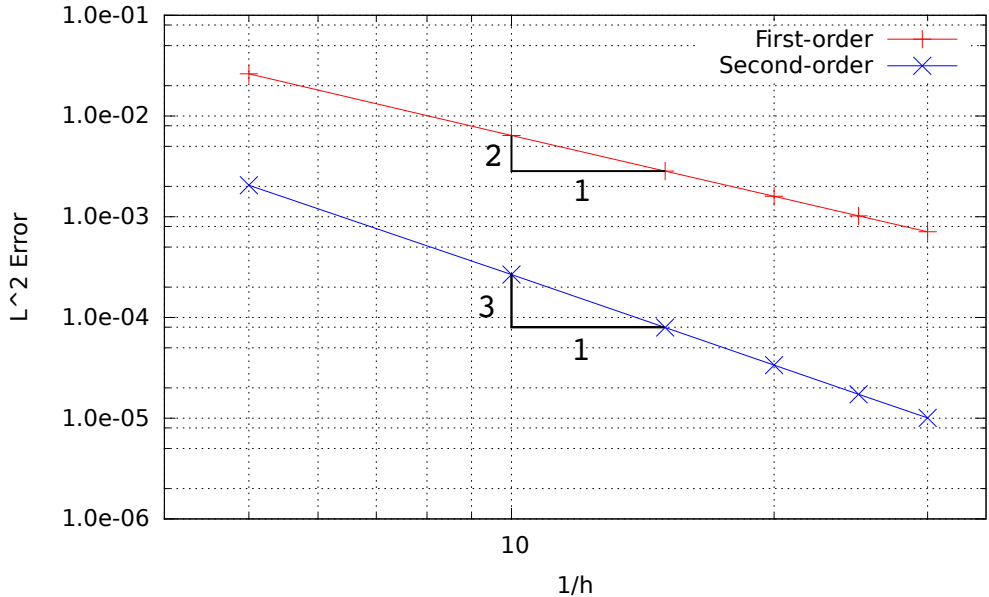}
\caption{Spatial convergence plot (dt = 1e-4s). Convegence rates for first- and second-order hexahedral  elements.}
\label{figure.spatial_convergence}
\end{figure}
\begin{table}
  \centering
  \begin{tabular}{c|c|c|c|c|}
    \cline{2-5}
    &\multicolumn{2}{|c|}{P1} & \multicolumn{2}{|c|}{P2}\\
    \hline
    \multicolumn{1}{|c|}{$1/h$} & Error & Order & Error & Order\\
    \hline
    \multicolumn{1}{|c|}{$5$} & $2.63e{-2}$ & $-$ & $2.05e{-3}$ & $-$\\
    \multicolumn{1}{|c|}{$10$} & $6.42e{-3}$ & $2.04$ & $2.67e{-4}$ & $2.94$\\
    \multicolumn{1}{|c|}{$15$} & $2.84e{-3}$ & $2.01$ & $7.96e{-5}$ & $2.98$ \\
    \multicolumn{1}{|c|}{$20$} & $1.60e{-3}$ & $2.01$ & $3.37e{-5}$ & $2.99$ \\
    \multicolumn{1}{|c|}{$25$} & $1.02e{-3}$ & $2.00$ & $1.73e{-5}$ & $2.99$ \\
    \multicolumn{1}{|c|}{$30$} & $7.08e{-4}$ & $2.00$ & $1.01e{-5}$ & $2.96$ \\
    \hline
  \end{tabular}
  \caption{Error table of the spatial convergence study.}
  \label{table.space_convergence}
\end{table}
\figref{figure.spatial_convergence} and \tabref{table.space_convergence} show that the expected spatial convergence rate of the order $p+1$, where $p$ is the order of the polynomial interpolation, is indeed achieved.

\subsubsection{Temporal convergence}
\begin{figure}[ht]
\centering
\includegraphics[width=0.95\textwidth]{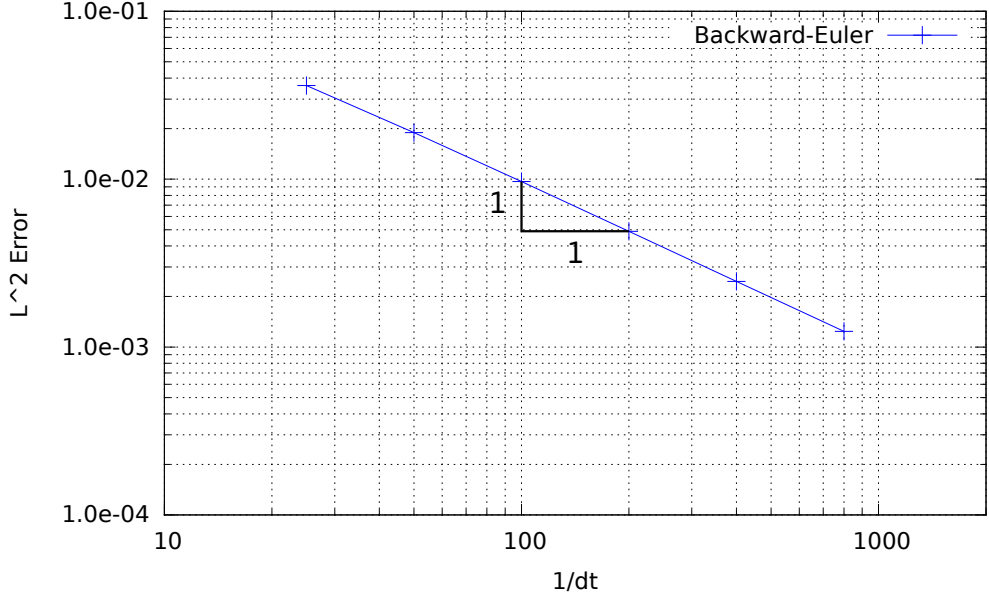}
\caption{Temporal convergence plot (h=1/15m). $L^2$-error of backward-Euler time-integration scheme.}
\label{figure.time_convergence}
\end{figure}

In this convergence study, the spatial discretization is assumed constant and fine enough such that the spatial discretization error is negligible. Second-order hexahedral elements, with a characteristic element size of $h = 1/15$, are used. \figref{figure.time_convergence} shows the expected first-order convergence rate, for a backward-Euler time-integration scheme. In \tabref{table.time_convergence}, the computed errors and convergence orders are given.

\begin{table}
  \centering
  \begin{tabular}{c|c|c|}
    %\cline{2-3}
    %&\multicolumn{2}{|c|}{P2}\\
    \hline
    \multicolumn{1}{|c|}{$1/\Delta t$} & Error & Order \\
    \hline
    \multicolumn{1}{|c|}{$25$} & $3.61e{-2}$ & $-$\\
    \multicolumn{1}{|c|}{$50$} & $1.89e{-2}$ & $0.93$\\
    \multicolumn{1}{|c|}{$100$} & $9.67e{-3}$ & $0.97$\\
    \multicolumn{1}{|c|}{$200$} & $4.89e{-3}$ & $0.98$\\
    \multicolumn{1}{|c|}{$400$} & $2.46e{-3}$ & $0.99$\\
    \multicolumn{1}{|c|}{$800$} & $1.24e{-3}$ & $0.99$\\
    \hline
  \end{tabular}
  \caption{Error table of the time convergence study.}
  \label{table.time_convergence}
\end{table}
\subsection{Wave propagation}
\label{sec.wave_propagation}
In this section, the propagation of a plane wave in free space is considered. The simulation domain is a spherical portion of free space with radius $R = 1.5\,$m. The domain is discretised with about $40000$ tetrahedral elements and ABCs as defined in \eqref{eq.abc} are used to bound the numerical domain. The total time for this simulation is $T_{\mathrm{max}} = 3.33\times10^{-8}\,$s, and the time-step length is $\Delta t = 10^{-11}\,$s. The vacuum electromagnetic properties are $\varepsilon_0=8.854\times10^{-12}\,\text{F/m} \text{ and } \mu_0=1.257\times10^{-6}\,\text{H/m}$.\par
The incoming wave propagates in the $y$-direction and has a frequency of $f = 300\,\text{MHz}$. The respective electric and magnetic fields are given as
\begin{equation}
  \label{eq.wave_inc}
  \begin{split}
  &\mathbf{E}^{\mathrm{inc}} = \left\{
  \begin{matrix*}[l]
    E^{\mathrm{inc}}_x = 0,\\[2mm]
    E^{\mathrm{inc}}_y = 0,\\[2mm]
    E^{\mathrm{inc}}_z = \cos(\sqrt{\varepsilon_0\mu_0}\,\omega y-\omega t),%\\[2mm]
  \end{matrix*}\right.\\
  &\mathbf{H}^{\mathrm{inc}} = \left\{
  \begin{matrix*}[l]
    H^{\mathrm{inc}}_x = \sqrt{\frac{\varepsilon_0}{\mu_0}}\cos(\sqrt{\varepsilon_0\mu_0}\,\omega y-\omega t),\\[2mm]
    H^{\mathrm{inc}}_y = 0,\\[2mm]
    H^{\mathrm{inc}}_z = 0,
  \end{matrix*}\right.
\end{split}
\end{equation}
where $\omega = 2\pi f$ is the angular frequency of the wave. The velocity of the wave is given by $c_0 = 1/\sqrt{\varepsilon_0\mu_0}$ and coincides with the physical speed of light in vacuum.\par
As in Section \ref{subsection.cubicpec}, it is possible to compute the error with respect to an analytical solution of the plane wave as given by \eqref{eq.wave_inc}. The results of this simulation can be used as a reference state for the scattering of a plane wave. The error in the $L_2$-norm of the electric field is $\|\mathbf{E}^{\mathrm{inc}} - \mathbf{E}^h\|_{\tau_h} = 0.16814$ at $t = 10^{-8}$s, corresponding to about three periods of oscillation of the wave. Note that $c_0 = 1/\sqrt{\mu_0\varepsilon_0} \approx 3\times10^{8}$ m/s, such that in $10^{-8}$ s the wave travels $3$ m, that is, the length of the domain.\par
To quantify the errors due to the spatial temporal discretizations, a second simulation is run, using a time-step length of $\Delta t_2 = 0.5\Delta t$. The error is computed at the same time $\|\mathbf{E}^{\mathrm{inc}} - \mathbf{E}^h\|_{\tau_h} = 0.15471$, indicating that the error is mainly due to the spatial approximation.\par
\begin{figure}
  \centering
  \includegraphics[width=.8\textwidth]{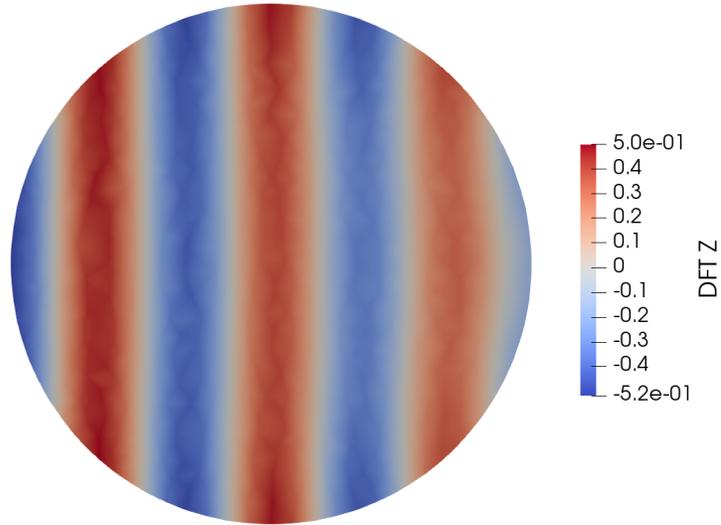}
  \caption{Wave propagation: DFT of $E_z$ component for $f=300\text{MHz}$.}
  \label{fig.wave_propagation_dft}
\end{figure}
In \figref{fig.wave_propagation_dft}, the real part of the Discrete Fourier Transform (DFT) of the solution is shown for the frequency $f = 300\,\text{MHz}$, allowing for a comparison with results in the frequency domain. The Fourier transform is computed by a pointwise temporal transformation using Numpy's FFT implementation (see, e.g., \cite{numpy}) and normalized by $1/n$, where $n$ is the number of time-steps used in the DFT. The normalization is required for enabling a comparison of the numerical results to the analytical Fourier transform of the plane wave, which is given by
\begin{equation}
  \mathbf{\hat{E}}^{\mathrm{inc}}(\mathbf{x}, f) = \frac{1}{2}\cos\left(\sqrt{\varepsilon_0\mu_0}2\pi f_0 \mathbf{x}\right)\delta(f_0-f),
\end{equation}
where $f_0$ is the frequency of the wave and $\delta(f_0 - f)$ the Dirac delta function. The Fourier transform of $E_z$ for the frequency $f_0 = 300$ MHz reads
\begin{equation}
  \hat{E}_z^{\mathrm{inc}}(\mathbf{x}) = \frac{1}{2}\cos\left(\sqrt{\varepsilon_0\mu_0}2\pi f_0 y\right),
\end{equation}
and it can be observed that the Fourier transform of the electric field is related to the electric field itself by the relation
\begin{equation}
  \mathbf{\hat{E}}^{\mathrm{inc}}(\mathbf{x}, f=300\text{MHz}) = \frac{1}{2}\mathbf{E}^{\mathrm{inc}}(\mathbf{x}, t=0).
\end{equation}
The numerical error in the $L_2$-norm of $\hat{E}_z^h$ is $\|\hat{E}^{\mathrm{inc}}_z - \hat{E}_z^h\|_{\tau_h} = 0.2763$.\par
Note that, in the simulations, the incoming wave is partially reflected by the boundaries, reducing the accuracy of the solution especially in the region of the positive $y$-axis.

\subsection{Wave scattering}
\label{sec.wave_scattering}
\begin{figure}
  \centering
  \includegraphics[width=.5\textwidth]{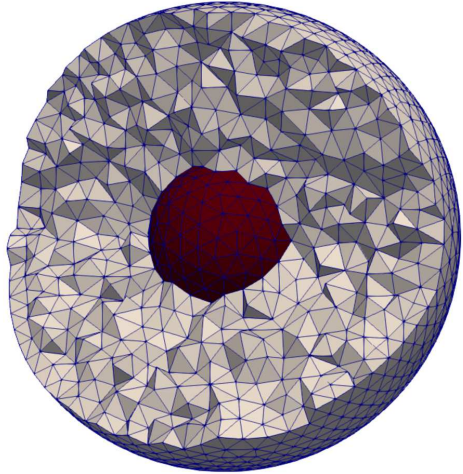}
  \caption{Computational domain for wave-scattering simulation, with scattering caused by the dielectric sphere in center of domain.}\label{fig.scatteringdomain}
\end{figure}

In this test case, the scattering of a plane wave by means of a dielectric sphere is computed. The domain $\Omega_1$ is spherical, with radius $R = 1.5\,\text{m}$, and contains a sphere $\Omega_2$ of radius $r = 0.5\,\text{m}$ and dielectric material, as shown in \figref{fig.scatteringdomain}.
On the boundary of $\Omega_1$, ABCs of Silver--M\"uller type with the components of the wave defined in \eqref{eq.wave_inc} are prescribed. The spatial discretization is the same as in Section \ref{sec.wave_propagation}. The material properties in $\Omega_1$ are also the same as in \ref{sec.wave_propagation}, while in $\Omega_2$, it is assumed that $\varepsilon_2 = 2\,\varepsilon_0$ and $\mu_2 = \mu_0$.\par
\begin{figure}\centering
  \includegraphics[width=.49\textwidth]{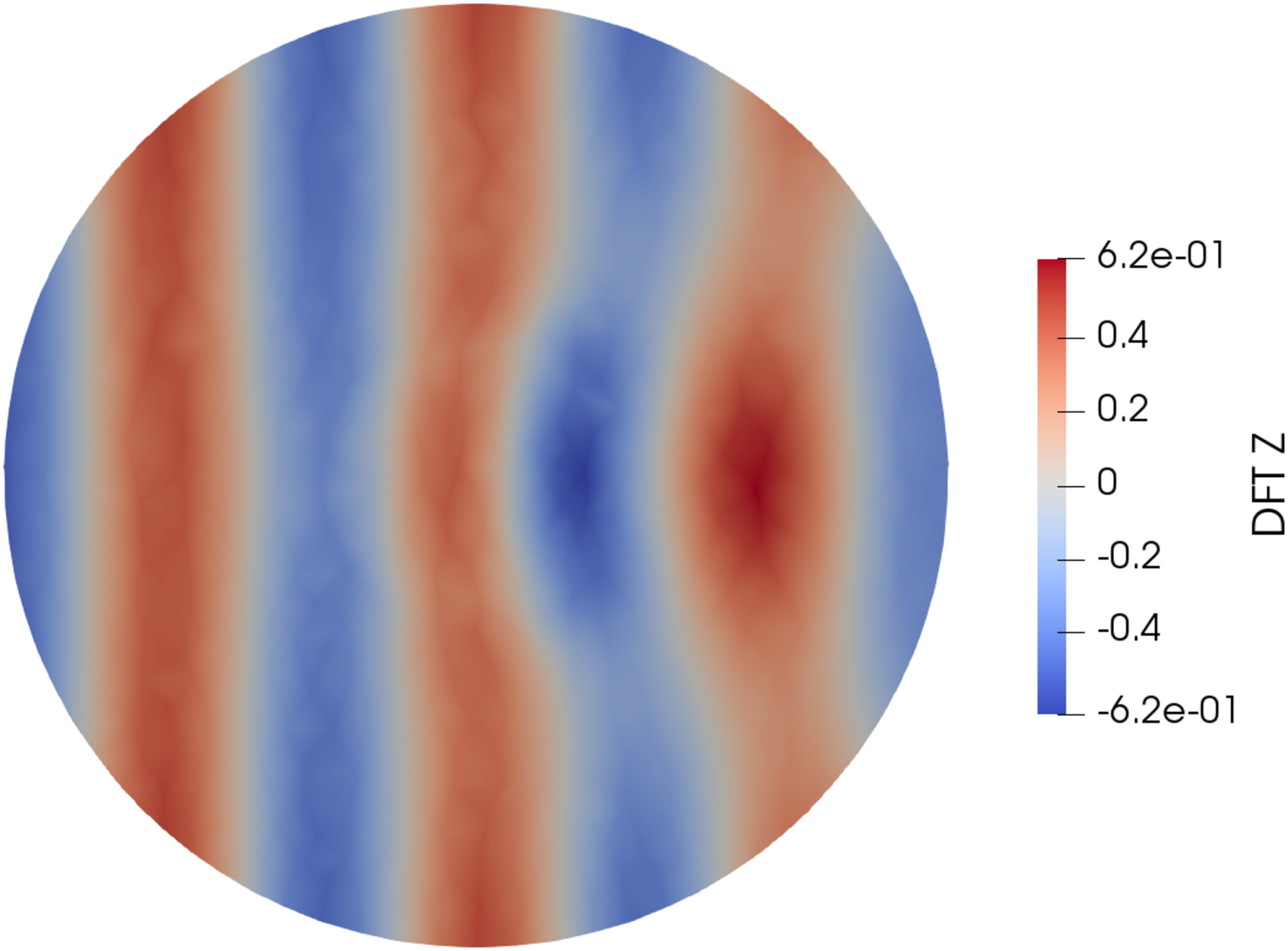}
  \includegraphics[width=.49\textwidth]{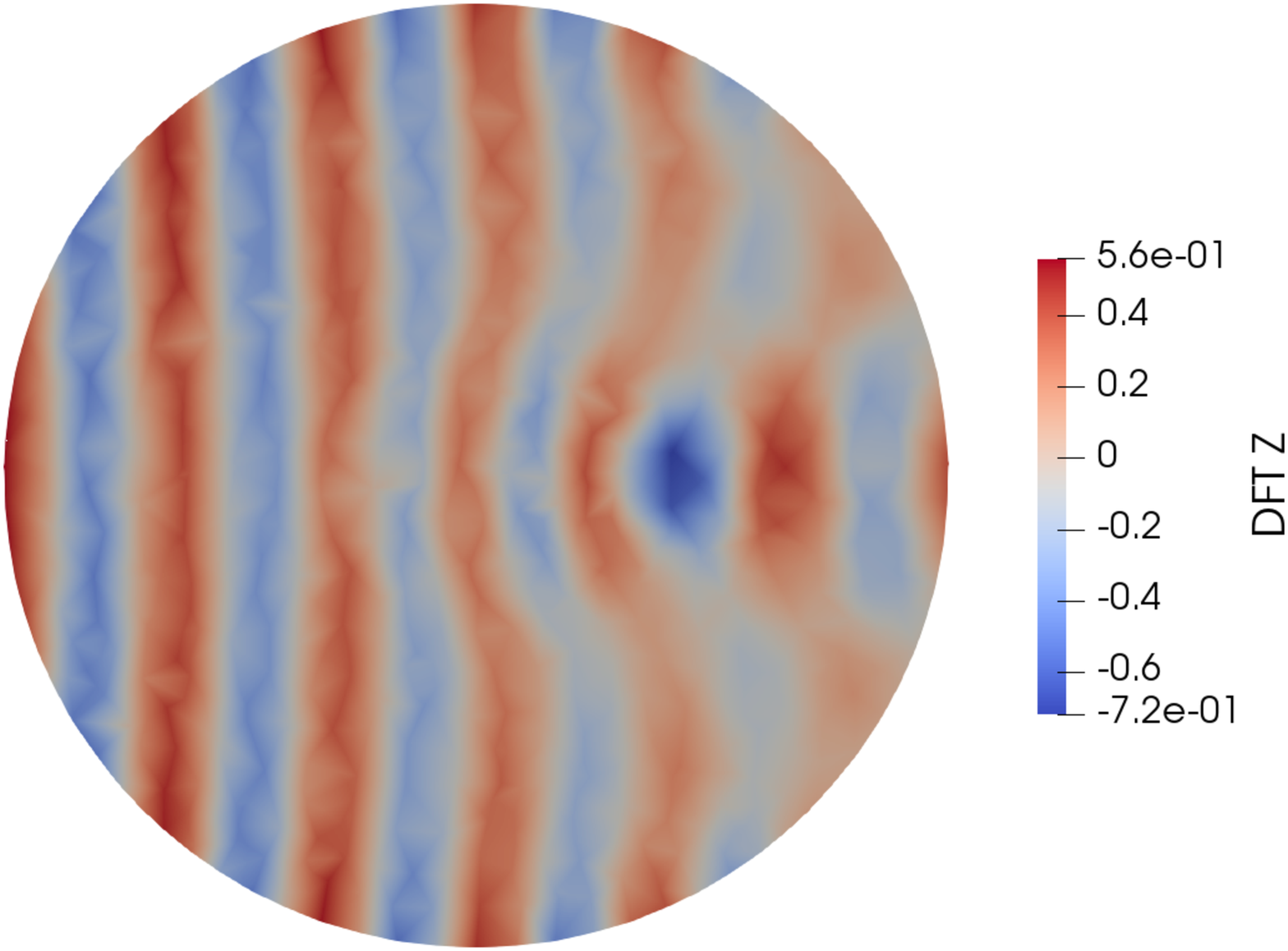}
  \caption{Wave scattering: DFT of $E_z$ component, $f=300\text{MHz}$ on the left-hand side and $f=600\text{MHz}$ on the right-hand side.}
  \label{fig.wave_scattering}
\end{figure}

For this test case, two different simulations frequencies are considered. \figref{fig.wave_scattering} shows the scattering of a wave oscillating with a frequency $f = 300\text{MHz}$ on the left-hand side and $f = 600\text{MHz}$ on the right-hand side. Again, For $f = 300\text{MHz}$, the time-integration parameters are the same as in Section \ref{sec.wave_propagation}, that is $T_{\mathrm{max}} = 3.33\times10^{-8}\,$s and $\Delta t = 10^{-11}\,$s. For $f = 600\text{MHz}$, due to the higher frequency of the wave, both simulation time $T_{\mathrm{max}}$, and time-step length $\Delta t$ are halved, such that $T_{\mathrm{max}} = 1.67\times10^{-8}\,$s and $\Delta t = 5\times10^{-12}\,$s.

\subsection{Cubic cavity with perfect electric conductive boundary conditions in conductive media}
To validate the formulation for problems in conductive environments, such as those that can be found in geophysical applications, a numerical example based on a manufactured solution is used. In this test case, permittivity and permeability values are set to $\varepsilon = \sqrt{3}\,\text{F/m}$ and $\mu = \sqrt{3}\,\text{H/m}$, respectively, and a non-zero conductivity value is introduced: $\sigma = 1.429$ S/m. This conductivity is inspired by values used in \cite{paper.um10} as exemplary conductivities of seabed models.\par
The manufactured solution aims at enforcing the analytical solution \eqref{eq.initial} in the mixed equations \eqref{mixeq}, yielding
\begin{equation}\label{eq.forc_term}
  \mathbf{i}_{\mathrm{s}} = - \sigma\mathbf{E}^{\mathrm{a}},
\end{equation}
as a constraint for the forcing term. As a result, if $\mathbf{i_{\mathrm{s}}}$ satisfies \eqref{eq.forc_term}, the analytical solution \eqref{eq.initial} will satisfy equations \eqref{mixeq}.\par
The time discretization is the same as in Section \ref{subsubsection.spatial_convergence}, that is, $\Delta t = 10^{-4}$ and $T_{max} = 5\times10^{-4}$. Different spatial discretizations are used to simulate the proposed problem. First, hexahedral elements within structured meshes are used. A spatial convergence study is carried out by discretising the domain with the meshes used in Section \ref{subsubsection.spatial_convergence}. The obtained error values and spatial convergence rates are identical to the ones shown in Table \ref{table.space_convergence}, indicating that the conductivity term is not critical in terms of the accuracy of the discretization scheme.\par
\begin{figure}\centering
  \includegraphics[width=.5\textwidth]{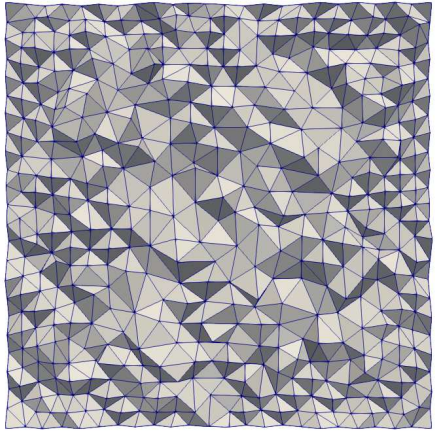}
  \caption{Midsection of unstructured mesh for the cubic cavity with conductive media (characteristic element size approximately $h = 1/20$).}\label{fig.tet_mesh}
\end{figure}

Additionally, the domain is dicretised using tetrahedral elements within an unstructured mesh, as exemplary depicted in \figref{fig.tet_mesh}.
A convergence study is carried out using increasingly finer meshes, and the results are reported in Table \ref{table.cubic_tet}. As expected, the convergence rate only approximates the convergence rate expected for linear elements, due to the non uniform element size across the domain.
\begin{table}
  \centering
  \begin{tabular}{c|c|c|}
    \cline{2-3}
    &\multicolumn{2}{|c|}{P1}\\
    \hline
    \multicolumn{1}{|c|}{$1/h$} & Error & Order \\
    \hline
    \multicolumn{1}{|c|}{$5$} & $3.34e{-2}$ & $-$\\
    \multicolumn{1}{|c|}{$10$} & $8.20e{-3}$ & $2.03$\\
    \multicolumn{1}{|c|}{$15$} & $4.40e{-3}$ & $1.84$\\
    \multicolumn{1}{|c|}{$20$} & $2.58e{-3}$ & $1.85$\\
    \hline
  \end{tabular}
  \caption{Error table of the spatial convergence study of a cubic cavity with conductive media.}
  \label{table.cubic_tet}
\end{table}

\section{Conclusions}
\label{section.conclusions}
Two implicit Hybridizable Discontinuous Galerkin formulations for the solution of Maxwell's equations in three-dimensional domains have been presented. The formulations are designed to be used in domains with diffusive mediums, and therefore, the conductivity term is present in both formulations. While the formulation for the electric field only takes into account the conduction current, the formulation for the mixed equations is more general in that it retains both displacement and conduction currents.\par
The formulation for the mixed equations has been validated for three well-known test cases, such as the resonance in a cubic cavity, wave propagation and scattering, and for a manufactured solution to particularly take into account the conductivity term. Spatial convergence studies have been carried out for the wave-resonance test case in dielectric media, and the expected convergence rates have been obtained. Solutions of wave propagation and scattering have been proposed in time and frequency domain, allowing for comparisons with different solution methods. The case of a resonant cavity with conductive media has also been simulated, and the accuracy of the results turned out to be identical to the test case with dielectric media.\par
The diffusive term in the equations will be particularly investigated in future work on applications in the field of geophysics, that is, for transient electromagnetics in three-dimensional diffusive earth media, as were already considered before, e.g., in \cite{paper.um10, paper.um12}. Future research will also focus on solvers and preconditioners for HDG methods for electromagnetics. Moreover, a comparison between implicit and explicit time-integration will be investigated. Numerical evidence shows that, even though explicit schemes can have severe time-step restrictions, explicit implementations may still be more efficient than implicit ones in the end; see, e.g., \cite{ksmw16}, where such a comparison was carried out for HDG methods in the context of problems of acoustic wave propogation, which exhibit similarities to the present problems.\par

\section*{Acknowledgements}
The support by the German Federal Ministry for Economic Affairs and Energy via the Central SME Innovation Program “Zentrales Innovationsprogramm Mittelstand (ZIM)” (Grant No. ZF4415001PO7) is gratefully acknowledged.



%% References

%\section*{References}

%\bibliographystyle{alpha}
\bibliography{elemag-hdg-paper_preprint}

\begin{thebibliography}{10}
\expandafter\ifx\csname url\endcsname\relax
  \def\url#1{\texttt{#1}}\fi
\expandafter\ifx\csname urlprefix\endcsname\relax\def\urlprefix{URL }\fi
\expandafter\ifx\csname href\endcsname\relax
  \def\href#1#2{#2} \def\path#1{#1}\fi

\bibitem{paper.yee}
K.~{Yee}, {Numerical solution of inital boundary value problems involving
  maxwell's equations in isotropic media}, IEEE Transactions on Antennas and
  Propagation 14 (1966) 302--307.
\newblock \href {http://dx.doi.org/10.1109/TAP.1966.1138693}
  {\path{doi:10.1109/TAP.1966.1138693}}.

\bibitem{paper.barucq}
H.~Barucq, F.~Delaurens, B.~Hanouzet,
  \href{http://www.jstor.org/stable/2587124}{{Method of Absorbing Boundary
  Conditions: Phenomena of Error Stabilization}}, SIAM Journal on Numerical
  Analysis 35~(3) (1998) 1113--1129.
\newline\urlprefix\url{http://www.jstor.org/stable/2587124}

\bibitem{paper.haber2004}
E.~Haber, U.~M. Ascher, D.~W. Oldenburg,
  \href{https://doi.org/10.1190/1.1801938}{{Inversion of 3D electromagnetic
  data in frequency and time domain using an inexact all‐at‐once
  approach}}, GEOPHYSICS 69~(5) (2004) 1216--1228.
\newblock \href {http://arxiv.org/abs/https://doi.org/10.1190/1.1801938}
  {\path{arXiv:https://doi.org/10.1190/1.1801938}}, \href
  {http://dx.doi.org/10.1190/1.1801938} {\path{doi:10.1190/1.1801938}}.
\newline\urlprefix\url{https://doi.org/10.1190/1.1801938}

\bibitem{paper.um10}
E.~S. Um, J.~M. Harris, D.~L. Alumbaugh,
  \href{https://doi.org/10.1190/1.3473694}{{3D time-domain simulation of
  electromagnetic diffusion phenomena: A finite-element electric-field
  approach}}, GEOPHYSICS 75~(4) (2010) F115--F126.
\newblock \href {http://arxiv.org/abs/https://doi.org/10.1190/1.3473694}
  {\path{arXiv:https://doi.org/10.1190/1.3473694}}, \href
  {http://dx.doi.org/10.1190/1.3473694} {\path{doi:10.1190/1.3473694}}.
\newline\urlprefix\url{https://doi.org/10.1190/1.3473694}

\bibitem{npc11b}
N.~Nguyen, J.~Peraire, B.~Cockburn,
  \href{http://www.sciencedirect.com/science/article/pii/S0021999111003226}{{Hybridizable
  discontinuous Galerkin methods for the time-harmonic Maxwell’s equations}},
  Journal of Computational Physics 230~(19) (2011) 7151 -- 7175.
\newblock \href {http://dx.doi.org/https://doi.org/10.1016/j.jcp.2011.05.018}
  {\path{doi:https://doi.org/10.1016/j.jcp.2011.05.018}}.
\newline\urlprefix\url{http://www.sciencedirect.com/science/article/pii/S0021999111003226}

\bibitem{cdl17}
A.~Christophe, S.~Descombes, S.~Lanteri,
  \href{http://www.sciencedirect.com/science/article/pii/S0096300317302758}{{An
  implicit hybridized discontinuous Galerkin method for the 3D time-domain
  Maxwell equations}}, Applied Mathematics and Computation 319 (2018) 395 --
  408, recent Advances in Computing.
\newblock \href {http://dx.doi.org/https://doi.org/10.1016/j.amc.2017.04.023}
  {\path{doi:https://doi.org/10.1016/j.amc.2017.04.023}}.
\newline\urlprefix\url{http://www.sciencedirect.com/science/article/pii/S0096300317302758}

\bibitem{Nedelec1980}
J.~C. N\'ed\'elec, \href{https://doi.org/10.1007/BF01396415}{{Mixed finite
  elements in R3}}, Numerische Mathematik 35~(3) (1980) 315--341.
\newblock \href {http://dx.doi.org/10.1007/BF01396415}
  {\path{doi:10.1007/BF01396415}}.
\newline\urlprefix\url{https://doi.org/10.1007/BF01396415}

\bibitem{Nedelec1986}
J.~C. N{\'e}d{\'e}lec, \href{https://doi.org/10.1007/BF01389668}{{A new family
  of mixed finite elements in R3}}, Numerische Mathematik 50~(1) (1986) 57--81.
\newblock \href {http://dx.doi.org/10.1007/BF01389668}
  {\path{doi:10.1007/BF01389668}}.
\newline\urlprefix\url{https://doi.org/10.1007/BF01389668}

\bibitem{paper.sun95}
D.~Sun, J.~Manges, X.~Yuan, Z.~Cendes, {Spurious modes in finite-element
  methods}, IEEE Antennas and Propagation Magazine 37~(5) (1995) 12--24.
\newblock \href {http://dx.doi.org/10.1109/74.475860}
  {\path{doi:10.1109/74.475860}}.

\bibitem{paper.unified_hybridization}
B.~Cockburn, J.~Gopalakrishnan, R.~Lazarov,
  \href{https://doi.org/10.1137/070706616}{{Unified Hybridization of
  Discontinuous Galerkin, Mixed, and Continuous Galerkin Methods for Second
  Order Elliptic Problems}}, SIAM Journal on Numerical Analysis 47~(2) (2009)
  1319--1365.
\newblock \href {http://arxiv.org/abs/https://doi.org/10.1137/070706616}
  {\path{arXiv:https://doi.org/10.1137/070706616}}, \href
  {http://dx.doi.org/10.1137/070706616} {\path{doi:10.1137/070706616}}.
\newline\urlprefix\url{https://doi.org/10.1137/070706616}

\bibitem{paper.yakovlev2016}
S.~Yakovlev, D.~Moxey, R.~M. Kirby, S.~J. Sherwin,
  \href{https://doi.org/10.1007/s10915-015-0076-6}{{To CG or to HDG: A
  Comparative Study in 3D}}, Journal of Scientific Computing 67~(1) (2016)
  192--220.
\newblock \href {http://dx.doi.org/10.1007/s10915-015-0076-6}
  {\path{doi:10.1007/s10915-015-0076-6}}.
\newline\urlprefix\url{https://doi.org/10.1007/s10915-015-0076-6}

\bibitem{paper.kronmatrixfree}
M.~Kronbichler, W.~Wall, \href{https://doi.org/10.1137/16M110455X}{A
  performance comparison of continuous and discontinuous galerkin methods with
  fast multigrid solvers}, SIAM Journal on Scientific Computing 40~(5) (2018)
  A3423--A3448.
\newblock \href {http://arxiv.org/abs/https://doi.org/10.1137/16M110455X}
  {\path{arXiv:https://doi.org/10.1137/16M110455X}}, \href
  {http://dx.doi.org/10.1137/16M110455X} {\path{doi:10.1137/16M110455X}}.
\newline\urlprefix\url{https://doi.org/10.1137/16M110455X}

\bibitem{paper.wang93}
T.~Wang, G.~W. Hohmann, \href{https://doi.org/10.1190/1.1443465}{{A
  finite-difference, time-‐domain solution for three-‐dimensional
  electromagnetic modeling}}, GEOPHYSICS 58~(6) (1993) 797--809.
\newblock \href {http://arxiv.org/abs/https://doi.org/10.1190/1.1443465}
  {\path{arXiv:https://doi.org/10.1190/1.1443465}}, \href
  {http://dx.doi.org/10.1190/1.1443465} {\path{doi:10.1190/1.1443465}}.
\newline\urlprefix\url{https://doi.org/10.1190/1.1443465}

\bibitem{paper.commer2004}
M.~Commer, G.~Newman, \href{https://doi.org/10.1190/1.1801936}{{A parallel
  finite‐difference approach for 3D transient electromagnetic modeling with
  galvanic sources}}, GEOPHYSICS 69~(5) (2004) 1192--1202.
\newblock \href {http://arxiv.org/abs/https://doi.org/10.1190/1.1801936}
  {\path{arXiv:https://doi.org/10.1190/1.1801936}}, \href
  {http://dx.doi.org/10.1190/1.1801936} {\path{doi:10.1190/1.1801936}}.
\newline\urlprefix\url{https://doi.org/10.1190/1.1801936}

\bibitem{paper.unsworth93}
M.~J. Unsworth, B.~J. Travis, A.~D. Chave,
  \href{https://doi.org/10.1190/1.1443406}{{Electromagnetic induction by a
  finite electric dipole source over a 2-D earth}}, GEOPHYSICS 58~(2) (1993)
  198--214.
\newblock \href {http://arxiv.org/abs/https://doi.org/10.1190/1.1443406}
  {\path{arXiv:https://doi.org/10.1190/1.1443406}}, \href
  {http://dx.doi.org/10.1190/1.1443406} {\path{doi:10.1190/1.1443406}}.
\newline\urlprefix\url{https://doi.org/10.1190/1.1443406}

\bibitem{paper.um12}
E.~Um, J.~Harris, D.~L.~Alumbaugh, {An iterative finite element time-domain
  method for simulating three-dimensional electromagnetic diffusion in earth},
  Geophysical Journal International 190 (2012) 871--886.
\newblock \href {http://dx.doi.org/10.1111/j.1365-246X.2012.05540.x}
  {\path{doi:10.1111/j.1365-246X.2012.05540.x}}.

\bibitem{paper.um14}
E.~Um, M.~Commer, G.~A.~Newman, {Efficient pre-conditioned iterative solution
  strategies for the electromagnetic diffusion in the Earth: finite-element
  frequency-domain approach}, Geophysical Journal International 193 (2013)
  1460--1473.
\newblock \href {http://dx.doi.org/10.1093/gji/ggt071}
  {\path{doi:10.1093/gji/ggt071}}.

\bibitem{paper.um15}
E.~Um, M.~Commer, G.~A.~Newman, G.~Hoversten, {Finite element modelling of
  transient electromagnetic fields near steel-cased wells}, Geophysical Journal
  International 202 (2015) 901--913.
\newblock \href {http://dx.doi.org/10.1093/gji/ggv193}
  {\path{doi:10.1093/gji/ggv193}}.

\bibitem{book.jackson99}
J.~D. Jackson, \href{http://cdsweb.cern.ch/record/490457}{{Classical
  electrodynamics}}, 3rd Edition, Wiley, New York, {NY}, 1999.
\newline\urlprefix\url{http://cdsweb.cern.ch/record/490457}

\bibitem{paper.rosen80}
J.~Rosen, \href{https://doi.org/10.1119/1.12289}{{Redundancy and superfluity
  for electromagnetic fields and potentials}}, American Journal of Physics
  48~(12) (1980) 1071--1073.
\newblock \href {http://arxiv.org/abs/https://doi.org/10.1119/1.12289}
  {\path{arXiv:https://doi.org/10.1119/1.12289}}, \href
  {http://dx.doi.org/10.1119/1.12289} {\path{doi:10.1119/1.12289}}.
\newline\urlprefix\url{https://doi.org/10.1119/1.12289}

\bibitem{paper.haber2000}
E.~Haber, \href{https://doi.org/10.1023/A:1011540222718}{{A mixed finite
  element method for the solution of the magnetostatic problem with highly
  discontinuous coefficients in 3D}}, Computational Geosciences 4~(4) (2000)
  323--336.
\newblock \href {http://dx.doi.org/10.1023/A:1011540222718}
  {\path{doi:10.1023/A:1011540222718}}.
\newline\urlprefix\url{https://doi.org/10.1023/A:1011540222718}

\bibitem{c10}
S.~Constable, \href{https://doi.org/10.1190/1.3483451}{{Ten years of marine
  CSEM for hydrocarbon exploration}}, GEOPHYSICS 75~(5) (2010) 75A67--75A81.
\newblock \href {http://arxiv.org/abs/https://doi.org/10.1190/1.3483451}
  {\path{arXiv:https://doi.org/10.1190/1.3483451}}, \href
  {http://dx.doi.org/10.1190/1.3483451} {\path{doi:10.1190/1.3483451}}.
\newline\urlprefix\url{https://doi.org/10.1190/1.3483451}

\bibitem{report.superlu99}
X.~Li, J.~Demmel, J.~Gilbert, iL. Grigori, M.~Shao, I.~Yamazaki, {SuperLU
  Users' Guide}, Tech. Rep. LBNL-44289, Lawrence Berkeley National Laboratory
  (September 1999).

\bibitem{paper.li03}
X.~S. Li, J.~W. Demmel, {SuperLU\_DIST}: A scalable distributed-memory sparse
  direct solver for unsymmetric linear systems, ACM Trans. Mathematical
  Software 29~(2) (2003) 110--140.

\bibitem{paper.heroux05}
M.~A. Heroux, R.~A. Bartlett, V.~E. Howle, R.~J. Hoekstra, J.~J. Hu, T.~G.
  Kolda, R.~B. Lehoucq, K.~R. Long, R.~P. Pawlowski, E.~T. Phipps, A.~G.
  Salinger, H.~K. Thornquist, R.~S. Tuminaro, J.~M. Willenbring, A.~Williams,
  K.~S. Stanley, \href{http://doi.acm.org/10.1145/1089014.1089021}{{An Overview
  of the Trilinos Project}}, ACM Trans. Math. Softw. 31~(3) (2005) 397--423.
\newblock \href {http://dx.doi.org/10.1145/1089014.1089021}
  {\path{doi:10.1145/1089014.1089021}}.
\newline\urlprefix\url{http://doi.acm.org/10.1145/1089014.1089021}

\bibitem{numpy}
T.~Oliphant, \href{http://www.numpy.org/}{{NumPy}: A guide to {NumPy}}, USA:
  Trelgol Publishing (2006).
\newline\urlprefix\url{http://www.numpy.org/}

\bibitem{ksmw16}
{Kronbichler, M. and Schoeder, S. and M\"uller, C. and Wall, W. A.},
  \href{https://onlinelibrary.wiley.com/doi/abs/10.1002/nme.5137}{{Comparison
  of implicit and explicit hybridizable discontinuous Galerkin methods for the
  acoustic wave equation}}, International Journal for Numerical Methods in
  Engineering 106~(9) (2016) 712--739.
\newblock \href
  {http://arxiv.org/abs/https://onlinelibrary.wiley.com/doi/pdf/10.1002/nme.5137}
  {\path{arXiv:https://onlinelibrary.wiley.com/doi/pdf/10.1002/nme.5137}},
  \href {http://dx.doi.org/10.1002/nme.5137} {\path{doi:10.1002/nme.5137}}.
\newline\urlprefix\url{https://onlinelibrary.wiley.com/doi/abs/10.1002/nme.5137}

\end{thebibliography}
%\bibliographystyle{elsarticle-num}






% \end{thebibliography}

\end{document}